\documentclass[12pt, a4paper]{scrartcl}
\usepackage{lmodern}

\usepackage{amsmath, empheq, amsfonts, amssymb, amsthm, color, graphicx,
            subfigure, paralist} %
\usepackage[bookmarks,colorlinks,breaklinks]{hyperref}
\definecolor{dullmagenta}{rgb}{0.4,0,0.4}   
\definecolor{darkblue}{rgb}{0,0,0.4}
\definecolor{darkgreen}{rgb}{0,0.6,0}
\definecolor{darkred}{rgb}{0.6,0,0}
\hypersetup{linkcolor=darkblue,citecolor=blue,filecolor=dullmagenta,urlcolor=darkblue} 
	\newcommand{\ncd}{\newcommand}
	\ncd{\mrm}    {\mathrm}
	\ncd{\beq} {\begin{equation}}
	\ncd{\eeq} {\end{equation}}
	\def\w{{w}}

	\def\d{{\rm d}}

\begin{document}
\title{Contact Symmetries and Hamiltonian Thermodynamics}
\author{A Bravetti\thanks{
		Instituto de Ciencias Nucleares, Universidad Nacional Autonoma de M\'exico,
		A.P. 70-543, 04510 Mexico D.F., Mexico, \texttt{bravetti@correo.nucleares.unam.mx}
	}\,
	C S Lopez-Monsalvo\thanks{
		Instituto de Ciencias Nucleares, Universidad Nacional Autonoma de M\'exico,
		A.P. 70-543, 04510 Mexico D.F., Mexico, \texttt{cesar.slm@correo.nucleares.unam.mx}
	}\,
	F Nettel\thanks{
		Dipartimento di Fisica, Universit\`a di Roma La Sapienza, P.le Aldo Moro 5, I-00185 Rome, Italy,
		 \texttt{Francisco.Nettel@roma1.infn.it}
	}
\date{}
}

\date{\today}

\maketitle





	\begin{abstract}
It has been shown that contact geometry is the proper framework underlying classical thermodynamics and that thermodynamic fluctuations are captured by an additional metric structure related to Fisher's Information Matrix. In this work we analyze several unaddressed aspects about the application of contact and metric geometry to thermodynamics. We consider here the Thermodynamic Phase Space and start by investigating the role of gauge transformations and Legendre symmetries for metric contact manifolds and their significance in thermodynamics. Then we present a novel mathematical characterization of first order phase transitions as equilibrium processes on the Thermodynamic Phase Space for which the Legendre symmetry is broken. Moreover, we use contact Hamiltonian dynamics to represent thermodynamic processes in a way that resembles the classical Hamiltonian formulation of conservative mechanics and we show that the relevant Hamiltonian coincides with the irreversible entropy production along thermodynamic processes. Therefore, we use such property to give a geometric definition of thermodynamically admissible fluctuations according to the Second Law of thermodynamics. Finally, we show that the length of a curve describing a thermodynamic process measures its entropy production.
	\end{abstract}


\newpage\tableofcontents



\section{Introduction}

Several programmes for the geometrization of equilibrium thermodynamics {and thermodynamic fluctuation theory} have been proposed so far and the literature on the subject is vast 
{(see e.g. \cite{Gibbs} for the original works of Gibbs, \cite{wein1975,rupp1979,rupp1995,rupp2010,rupp2012} for the introduction
of Riemannian geometry based on Hessian structures on the equilibrium manifold,
\cite{Hermann,mrugala1,mrugala2} for the construction of the contact phase space of thermodynamics and 
\cite{SalamonBerryPRL,Schlogl,Montesinos,GTD,BrodyRivier,Crooks,CrooksPRE2012,CrooksPRL2012,CRGTD,iofra,LiuLu,MNSS1990} for later developments of all those aspects).}
In particular, 
it has been {shown} that the relevant manifolds are: 
 the Thermodynamic Phase Space (TPS) -- which is a contac manifold -- together with its Legendre sub-manifolds representing the spaces of equilibrium states associated to particular systems.
 We refer to \cite{mrugala1,mrugala2} and \cite{GTD} for the definitions of these two manifolds and the description of the 
 mappings between them. 
 Moreover, to incorporate thermodynamic fluctuations out of the equilibrium values into the construction, one equips the TPS with the (pseudo-)Riemannian
 structure stemming from a statistical moment expansion of the underlying micro-physics. 
 This in turn endows each space of equilibrium states with a metric structure induced from that of the TPS.

{The geometry of the equilibrium manifold has been largely studied providing us with
a number of interesting results about the thermodynamics of ordinary systems. In  particular, it has been proved that
the thermodynamic curvature diverges at critical points with the same critical exponent as the correlation volume \cite{rupp1995}. 
It has also been established that methods of thermodynamic geometry can help to find the optimum protocols
as well as the available work dissipated in the context of non-equilibrium thermodynamics \cite{SalamonBerryPRL,Crooks,CrooksPRE2012,CrooksPRL2012}. 
Furhtermore, in the thermodynamics of black holes,  it has been shown that the appearance of instabilities is related to the divergences 
of different Hessian structures on the equilibrium manifold \cite{iofra,LiuLu}.

 The contact metric geometry of the TPS, however, has received much less attention (except for 
 some initial  works by Mrugala and his collaborators \cite{mrugala1,mrugala2,MNSS1990} and some recent connection
with quantization due to Rajeev \cite{rajeevquantization}).}
 

{The aim of the present work is to widen the point of view on thermodynamic geometry and consider all the thermodynamic structures from
a more general perspective, that is, the contact metric geometry of the TPS.
 So far} it has been shown {(see \cite{MNSS1990} and \cite{TPSSASAKI})}  that both the contact and the metric 
structures on the TPS can be derived from statistical mechanics and information theory.
In particular, the contact structure of the TPS is generated by a 1-form emerging from  the first variation of the relative entropy near Gibbs' equilibrium distribution, 
providing us with a mathematical strucutre encoding the First Law, 
while the 
(pseudo-)Riemannian metric, obtained from the second variation, represents thermodynamic fluctuations \cite{TPSSASAKI}. 
Furthermore, such (pseudo-)Riemannian structure induces  the well-known Fisher-Rao 
Information Metric  into the Legendre sub-manifolds of the TPS \cite{BrodyRivier,Crooks}.   
Finally, a further exploration of the geometric properties of the TPS reveals that it is 
a {para-Sasakian} and $\eta$-Einstein manifold \cite{SASAKI, Yano,libroBlair,Takahashi,Zamkovoy2009,IVZ,Perrone}, 
{which is
locally isomorphic to the Hyperbolic Heisenberg Group  \cite{TPSSASAKI}}. This construction is appealing due to its potential  connections with other branches of theoretical physics.
{For example, the emergence of the Heisenberg Group indicates some physical links with the quantum uncertainty relations in line with the analysis of \cite{rajeevquantization}. 
Therefore, from our point of view, it is worth to studying the mathematical symmetries of the TPS in order to exploit their physical significance.}


This work is meant to be a comprehensive presentation of several {uninvestigated} topics in the contact metric description
of thermodynamics. Therefore it can be divided into two parts. In the first part -- sections \ref{sec:gaugeandlegendre} and \ref{sec:Lsym} -- we
derive new results on the application of contact transformations to the para-Sasakian structure of the TPS.
In the second part -- section \ref{sec:ECH} -- we present our main result, that is, a contact Hamiltonian formulation
for `thermo-dynamics'. Readers interested only in the latter result can,  in principle, skip sections \ref{sec:gaugeandlegendre} and \ref{sec:Lsym}.

{In the first part}, we consider two types of well-known transformations of the contact metric structure of the TPS:
the re-scalings  of the 1-form generating the same contract structure (or \emph{gauge transformations})
and those transformations leaving the 1-form unchanged (or \emph{strict contactomorphisms or contact symmetries}).
In this work, we focus our attention  on the second type of transformations, i.e. contact symmetries, leaving the discussion
about the physical interpretation of gauge transformations to future work \cite{usinpreparation}. Here, we simply highlight that a change in the thermodynamic 
representation is a particular example of a gauge transformation inducing the well-known conformal
scaling between Weinhold's and Ruppeiner's metrics on the equilibrium manifolds \cite{SalamonRW}.

By investigating contact symmetries, we present Legendre transformations as an example of this class.
We then show that the \emph{Legendre symmetry}   
is a necessary and sufficient condition in order to 
reduce the thermodynamic degrees of freedom from the $2 n+1$ variables of the TPS to the $n$ degrees of freedom of the sub-manifold of equilibrium states.
However, when passing from the TPS to the equilibrium manifold representing a particular thermodynamic system, the Legendre symmetry can break.
In fact, as we will show, there is not just one sub-manifold of equilibrium, but  there are in principle as many as different statistical ensembles
(this was already observed in \cite{iofra,LiuLu}).
When the Legendre symmetry is valid, the different ensembles are equivalent and the change from one ensemble to the other is interpreted geometrically simply as a diffeomorphism 
between the 
equilibrium manifolds.
However, when the Legendre symmetry breaks, the ensembles are inequivalent and geometrically the map from one equilibrium sub-manifold to the other is not a diffeomorphism.
In this case the full description of the system cannot be achieved by means of a single ensemble. Therefore, we need to consider the full TPS instead of just one of its equilibrium sub-manifolds. In ordinary thermodynamics this happens only 
 at first order phase transitions. Thus we give a geometric characterization of first order phase transitions 
 as equilibrium processes on the full Thermodynamic Phase Space for which the Legendre symmetry is broken. 
This is the first result of the present work. 

Notice that if one also considers fluctuations, these are described differently in the distinct ensembles. Therefore in general the Legendre symmetry need not be also
an isometry between the different metric structures induced on the equilibrium sub-manifolds. However, we will show that a total Legendre transformation
always induces an isometry. This is the second result of this work.

In the second part of the manuscript, we consider a contact Hamiltonian formulation for thermodynamics, generalizing the work in \cite{mrugala2000,mrugalachina,mrugalaJPA}. 
In particular, we show that the irreversible entropy production over a fluctuation 
is captured by a  contact Hamiltonian on the TPS, thus in order to describe thermalization processes it is also necessary to resort to the TPS. 
Moreover, we prove that the corresponding contact Hamiltonian system is completely integrable, in a precise sense defined in \cite{boyerintegrablesystems}.
It turns out that the integral curves of this system define
thermodynamic processes at equilibrium or near equilibrium.
With this definition and based on the Second Law of thermodynamics we give
a simple characterization of thermodynamically admissible paths on the TPS. 
Finally, we show that using the  
metric structure on the TPS one can compute the entropy production along any admissible process.
We conclude by stating that this contact-Hamiltonian formulation of thermodynamic processes is morally tantamount to the
axiomatic version of the Laws of thermodynamics.
This is the third and main result presented in this work.

The outline of the paper is as follows. In Sec.~\ref{sec:Heisenberg} we review briefly the construction of the {contact and metric structures in the TPS from information geometry}.
In Sec.~\ref{sec:gaugeandlegendre} we discuss the role of gauge and Legendre transformations  in the geometry of thermodynamics.
In particular, we   argue that the first kind of transformations is not a symmetry of the contact bundle, while the second type is.
In Sec.~\ref{sec:Lsym} we   study in more detail the physical consequences of the Legendre symmetry of the Thermodynamic Phase Space and show that the breaking
of such symmetry for particular systems implies that there can be intersections between the equilibrium sub-manifolds and inequivalence of the ensembles. 
As a special case from ordinary thermodynamics, we   present the example of first order phase transitions.
In Sec.~\ref{sec:ECH} we   derive a contact Hamiltonian formulation of thermodynamics. We show that thermodynamic processes can be represented by
a completely integrable contact
Hamiltonian system where the relevant Hamiltonian is the irreversible entropy production.
Finally, we show that the length of any process equals its entropy production. 
 We conclude in Sec.~\ref{sec:conclusions} with a review of our results and a discussion about possible future investigations. 

\section{Contact and metric structures in Thermodynamics}\label{sec:Heisenberg}

In this section we provide the minimal geometric set up that will be relevant for the rest of this manuscript. 
Here, we present an account of some results on contact metric structures and their Legendre sub-manifolds in the light of a geometrization programme for thermodynamics.  
First, we {review the basics of Gibbs' statistical mechanics, focusing on the construction of different ensembles and we fix the notation.}
Then we
introduce the notion of a contact structure and the geometric objects associated to it. 
{Afterwards}, we present the link between these {geometric} objects and the statistical derivation of equilibrium thermodynamics. 
Finally, we present the Legendre sub-manifolds of the contact distribution as the constrained hypersurfaces defined by means of the First Law of thermodynamics.  

\subsection{Gibbs' distribution and statistical ensembles}\label{GibbsSM}

Let us start with the microscopic phase space of statistical mechanics $\Gamma$ whose volume measure is given by means of a \emph{normalizable} distribution $\rho$.
Define the microscopic entropy of $\rho$ as
	\beq\label{micros}
	s=-\text{ln}\rho
	\eeq 
and introduce Gibbs' entropy functional
	\beq
	\label{defentropy}
	S_{\rho} =\langle s(\rho) \rangle = -{\int_\Gamma \rho\, \text{ln}\rho\,\d \Gamma}.
	\eeq
Maximizing \eqref{defentropy} subject to the $n$ observational constraints 
	\beq  \label{constraint2}
	 p_i = \langle F_{i}\rangle =\frac{ \int \rho F^i\, \d\Gamma}{ \int \rho\, \d\Gamma}, \quad \text{with} \quad i=1,\dots, n,
	 \eeq
and the normalization condition
	\beq  \label{constraint1}
	\int  \rho\, \d\Gamma =1,
	\eeq 
one obtains the family of Gibbs' equilibrium distributions
	\beq\label{GibbsDistr}
	\rho_{0}(\Gamma; w,q^{1},\dots,q^{n})={\rm e}^{-w-F_{i}q^{i}},
	\eeq
where $q^{i}$ and $w$ correspond to the Lagrange multipliers of \eqref{constraint2} and \eqref{constraint1} respectively.

A direct calculation from equations \eqref{defentropy}, \eqref{constraint2} and \eqref{GibbsDistr} shows that the entropy for a distribution in the Gibbs' family  \eqref{GibbsDistr}  reads
	\beq\label{macroEnt}
	S_{\rho_{0}}=w+p_{i}q^{i},
	\eeq
and, therefore, $w$ turns out to be a Legendre transform of the entropy,
{whose significance changes according to the number and type of constraints in \eqref{constraint2}, i.e. depending on the 
statistical ensemble, as we will now discuss.

Let us consider simple systems -- those with only one species of particles -- in contact with different reservoirs. 
For an isolated system the internal energy $U$, the volume $V$ and the number of particles are all fixed and therefore none of them is allowed to fluctuate.
The corresponding ensemble is the microcanonical (or NVU) ensemble and the associated potential is the entropy $w(U,V,N)=S(U,V,N)$.
If the system is in contact only with a thermostat at temperature $T$, 
then the internal energy is the only variable which is allowed to fluctuate and $p_{1}=\langle H \rangle_{\rho_{0}}$
is the average value of the Hamiltonian energy, i.e. the internal energy $U$ of the system. In this case the number of particles $N$, the volume $V$ and 
the temperature $T$ are fixed. The corresponding ensemble is the canonical one (or NVT ensemble) 
and the associated thermodynamic potential is $w(N,V,T)=S-\beta U=-\beta F(N,V,T)$,
with $F(N,V,T)$ the Helmholtz free energy and $q^{1}=\beta=1/T$ the inverse temperature.

As another standard example, when the system is placed in contact both with a thermostat at temperature $T$ and a bariostat at pressure $p$, 
then energy and volume are the fluctuating variables.
In this case $p_{1}=\langle H \rangle_{\rho_{0}}$ as before, and  $p_{2}=\langle v \rangle_{\rho_{0}}$ 
is the average value of the volume of the system. Here $N, p$ and $T$ are fixed.
In this case the corresponding ensemble is the isothermal-isobaric ensemble (or NpT ensemble) and 
the associated potential is $w(N,T,p)=S-\beta U - \beta p V=-\beta G(N,T,p)$, where G(N,T,p) is the Gibbs free energy.

Finally, in the grand-canonical (or $\mu$VT) ensemble, the fluctuating variables are $p_{1}=\langle H \rangle_{\rho_{0}}$ and $p_{2}=\langle n \rangle_{\rho_{0}}$, while the fixed ones 
are $\mu,V$ and $T$. The associated thermodynamic potential is $w(\mu,V,T)=S-\beta U+\beta \mu N= - \beta \Phi(\mu,V,T) $, with $\Phi(\mu,V,T)$ the Grand potential, or Landau potential.

In principle one could also construct the \emph{`non-canonical ensemble'} (or $\mu$pT ensemble), i.e. an ensemble representing a system placed in 
contact with a thermostat, a bariostat and a particle reservoir. For such system all the extensive
variables $p_{1}=U, p_{2}=V$ and $p_{3}=N$ fluctuate, while the intensive ones $q^{1}=T, q^{2}=p$ and $q^{3}=\mu$ are fixed.
This situation is physically the easiest to realize. However, the associated thermodynamic potential $w(\mu,p,T)=S-\beta U-\beta p V+\beta\mu N$
identically vanishes, since the entropy $S$ is a homogeneous functions and therefore by Euler's relation $S=\beta U+\beta p V-\beta\mu N$. 
This is the reason why this ensemble is 
never considered in statistical mechanics. However, we will see in section \ref{sec:ECH} that this ensemble and its potential play a relevant role in our Hamiltonian description of thermodynamic
processes and fluctuations.

Notice that in all the above examples the constraints in \eqref{constraint2} define the internal variables allowed to fluctuate, while the corresponding 
Lagrange multipliers $q^{a}$ represent the external variables, whose values are fixed by the corresponding values of the reservoir.
Notice also that we take the variables $p_{a}$ to be the extensive variables of the system and the $q^{a}$ as the intensive ones.
In this notation the First Law for the entropy reads
	\beq\label{entropy1law}
	\d S=q^{a}\d p_{a},
	\eeq
and therefore the First Law for the total Legendre transform of the entropy is 
	\beq\label{w1law}
	\d w =-p_{a}\d q^{a}=0,
	\eeq
where the last equality in \eqref{w1law} is the Gibbs-Duhem identity \cite{Callen}.
}

\subsection{Contact metric manifolds: the Phase Space of thermodynamics}

Let us  consider the contact description of the Thermodynamic Phase Space as given e.g. in \cite{mrugala1,mrugala2,CRGTD}.
Given a thermodynamic system with $n$ degrees of freedom, the {\it Thermodynamic Phase Space} (TPS) is the $(2n+1)$-dimensional  manifold $\mathcal{T}$, 
endowed with a contact structure, that is, a maximally non-integrable distribution $\mathcal D\subset T\mathcal{T}$ of co-dimension one hyperplanes.  
We can characterize such a distribution with the aid of a 1-form $\eta$ such that 
	\beq\label{xi}
	\mathcal D=\ker(\eta), 
	\eeq 
and the non-integrability condition
\begin{equation}\label{integraxx}
   \eta \wedge (\d \eta)^n \neq 0
\end{equation}
is fulfilled. Equation \eqref{integraxx} can be understood as the condition for a well defined volume form on the TPS. 
Additionally, it is always possible to find a set of local (Darboux) coordinates for $\mathcal{T}$ such that the 1-form $\eta$ can be written in the form
	\begin{equation}\label{1stform}
	\eta = \d w + p_a \d q^a,
	\end{equation}
where we have used Einstein's convention for repeated indices and  $a$ takes values from $1$ to $n$, the number of degrees of freedom. 
Note that at this level $\w$, $q^a$ and $p_a$ are coordinates for $\mathcal{T}$ whose thermodynamic significance is linked to the underlying statistical mechanics 
{and that we are using here a different sign convention in $\eta$ with respect to previous work \cite{TPSSASAKI}. This sign convention was motivated by \eqref{entropy1law}.}

In the present work, we focus on the contact structure of $\cal T$.
First, note that the contact 1-form $\eta$ is not unique. Indeed, any other 1-form defining the same family of hyperplanes, equation  \eqref{xi}, 
is necessarily  conformally equivalent to $\eta$,
 i.e. for any two 1-forms $\eta_1$ and $\eta_2$ in the same equivalence class $[\eta]$, one has
$ \eta_2 = \Omega \,\eta_1$
for some non-vanishing real function $\Omega$.

Let us consider a contact 1-form $\eta$ defining the contact distribution $\mathcal{D}$. Associated to $\eta$ there is always a  global vector field $\xi$ --
the \emph{Reeb vector field} -- defined uniquely by the two conditions
\beq\label{Reeb}
\eta(\xi)=1 \quad \text{and} \quad \d \eta (\xi,\cdot)=0\,.
\eeq
The Reeb vector field  generates a natural splitting of the tangent bundle, that is 
\beq\label{splitting}
T\mathcal T = L_{\xi}\oplus \mathcal{D}\, ,
\eeq
where $ L_{\xi}$ is the \emph{vertical} sub-space generated by $\xi$.  
{In \cite{TPSSASAKI} it was shown that  the non-coordinate basis}
	\beq
	\label{basisTT}
	\Big{\{}\xi,\hat{P}^{i},\hat{Q}_{i}\Big{\}}=\left\{\xi,\,\frac{\partial}{\partial p_{i}},\,p_{i}\frac{\partial}{\partial w}-\frac{\partial}{\partial q^{i}}\right\}, \qquad i=1,\dots,2n
	\eeq
is naturally adapted to the splitting \eqref{splitting} 
{and that the generators 
satisfy the commutation relations}
	\beq
	\label{halgebra}
	[\hat P^i,\hat Q_j] = \delta^i_{\ j} \xi, \quad [\xi,\hat Q_i] = 0 \quad \text{and} \quad [\xi, \hat P^i] = 0,
	\eeq
defining the Lie-algebra of the $n$th Heisenberg group, $\mathcal{H}_n$. For this reason, we call the set  \eqref{basisTT}  {\emph{the Heisenberg basis of}} $T\mathcal{T}$.

{Analogously to the almost complex structure of K\"ahler manifolds, associated to each $\eta$ there is a $(1,1)$ tensor field $\phi$ such that 	
	\beq
	L_\xi = \ker(\phi) \quad \text{and}\quad \mathcal{D} = {\rm Im}(\phi). 
	\eeq
If the tensor field $\phi$ satisfies the condition
	\beq
	\phi^2 = - I + \eta \otimes \xi \quad \text{[resp.} \quad \phi^2 = I - \eta \otimes \xi \text{]}
	\eeq
it is called an \emph{almost contact structure} \cite{libroBlair} (resp. an \emph{almost para-contact structure} \cite{Zamkovoy2009}).  
Thus,  since $\phi(\xi) = 0$,  the splitting \eqref{splitting} becomes
	\beq
	\label{splitting2}
	T\mathcal{T} = \ker(\phi)  \oplus {\rm Im}(\phi).
	\eeq
 The quadruple $(\mathcal{T},\eta,\xi,\phi)$ is called an \emph{almost contact} [resp. \emph{para-contact}] \emph{structure}. 
 In general the tensor field $\phi$ is not unique. However,  if  $\mathcal{T}$ is equipped with a metric tensor, there is a preferred way for choosing $\phi$ such that
	\beq\label{associatedG}
	G(\phi X,\phi Y)=\varepsilon \left[ G(X,Y) - \eta(X)\,\eta(Y) \right],
	\eeq
for any pair of vector fields $X,Y \in T\mathcal{T}$ and where $\varepsilon = \pm 1$. In this case we say that the metric $G$ is a \emph{compatible metric} of the almost contact ($\varepsilon = 1$) [resp. para-contact ($\varepsilon = -1$)] structure $\phi$. 
Compatible metrics make the splitting \eqref{splitting2} orthogonal. 
Additionally, if the metric also satisfies
	\beq\label{associatedG1}
	 \frac{1}{2}\d \eta(X,Y) = G(X, \Phi Y) \,,
	 \eeq
we say that $G$ is an \textit{associated metric} to the contact structure and the 4-tuple $(\mathcal T,\eta,\xi,\phi,G)$ is a \emph{contact} [resp. \emph{para-contact}] \emph{metric manifold}  \cite{libroBlair,Takahashi,Zamkovoy2009}.}

\subsubsection{The Phase Space of Thermodynamics and the Hyperbolic Heisenberg Group}

Now,  considering  thermodynamic fluctuation theory, the TPS is not just a contact manifold, but it also carries  an almost {para}-contact structure and an associated metric. 
The derivation follows from information theory in the following way (c.f. \cite{TPSSASAKI} for details). 
Following \cite{TPSSASAKI}, {from equations \eqref{micros} and \eqref{GibbsDistr}} one can 
construct an $(n+1)-$dimensional control manifold $\mathcal{C}$ embedded in $\mathcal{T}$ such that 
	\beq
	\Phi : \mathcal{C} \longrightarrow \mathcal{T},
	\eeq
	\beq
	\label{THsasaki}
	\Phi^* \eta = \langle \d s\rangle_{0}=\d \w+ p_i\,\d q^i\,
	\eeq
and
	\beq
	\label{FisherRaorhoG2}
	\Phi^* G_{\rm FR} = \langle (\d s)^2 \rangle_{0} = \langle \d s\rangle_{0} \otimes \langle \d s\rangle_{0} -\d q^i \overset{\rm s}{\otimes}\d p_i\,,
	\eeq
where $\langle\cdot\rangle_{0}$ represents the ensemble 
average with respect to the equilibrium distribution $\rho_{0}$, $\Phi$ is the embedding of the control manifold into the TPS and $G_{\rm FR}$ is the
{metric on the TPS which reduces to the Fisher-Rao metric \eqref{FisherRaorhoG2} over the control manifold $\mathcal{C}$}.  
Here we have used the symbol $\overset{s}{\otimes}$ to denote the symmetric tensor product
	\beq
	\d q^i \overset{\rm s}{\otimes} \d p_i \equiv \frac{1}{2} \left(\d q^i \otimes \d p_i + \d p_i \otimes \d q^i \right).
	\eeq
Thus, in the coordinate basis of $T\mathcal T$, the metric $G_{\rm FR}$  reads explicitly as
	\beq\label{GinTcoordinate}
	G_{\rm FR}=(\d \w+ p_i\,\d q^i) \otimes (\d \w+ p_j\,\d q^j)-\d q^i \overset{\rm s}{\otimes} \d p_i\,.
	\eeq
Note that $\eta$ is naturally connected with the first moment of $\d s$ and hence with the First Law by means of \eqref{THsasaki}, while $G_{\rm FR}$ is connected with the second
moment of $\d s$ and hence with thermodynamic fluctuations. In fact, it turns out that the metric induced by $G_{\rm FR}$ into the equilibrium sub-manifolds is the Hessian of the corresponding 
thermodynamic potential.  Therefore it coincides with Ruppeiner's thermodynamic metric or its Legendre transformed analogues, depending on 
 the constraints that are considered \cite{TPSSASAKI}.

{The metric \eqref{GinTcoordinate} has an $(n+1,n)$ signature
{and it turns out that an orthonormal (dual) basis is given by}
	\beq\label{orthdualbasis}
	\left\{\hat \theta^{(0)}, \hat \theta^{(i)}_{+},\hat \theta^{(i)}_{-} \right\} 
	\eeq
where
	\beq
	\hat \theta^{(0)} = \eta \quad \text{and} \quad \hat  \theta^{(i)}_{\pm} = \frac{\sqrt{p_i}}{2 p_i}\left[-  p_i \d q^i  \pm  \d p_i\right] \quad \text{(no sum over $i$)}.
	\eeq
Thus, in terms of the Heisenberg basis \eqref{basisTT}, the $n$ `time-like' directions are given by 
		\beq\label{eminus}
		\hat e_{(i)}^- = -G^{-1}_{\rm FR}\left[\hat \theta^{(i)}_{-}, \cdot\right] = \sqrt{p_i} \left[\frac{1}{p_{i}}\hat Q_{i}-\hat{P^{i}}\right] \quad \text{(no sum over $i$)},
		\eeq
while the $n+1$ `space-like' directions are
		\beq\label{eplus}
		\hat e_{(0)} = \xi \quad \text{and} \quad \hat e_{(i)}^+ = G^{-1}_{\rm FR}\left[\hat \theta^{(i)}_{+}, \cdot \right] =  \sqrt{p_i} \left[\frac{1}{p_{i}}\hat Q_{i}+\hat{P^{i}}\right] \quad \text{(no sum over $i$)}.
		\eeq	
We call the basis \eqref{eminus}-\eqref{eplus} the \emph{canonical basis} of the TPS \cite{TPSSASAKI}.
Notice that the vectors $\{e_i^{+}\pm e_i^{-}\}$ are eigenvectors of the almost para-contact structure with eigenvalues $+1$ and $-1$, respectively.
	
The isometry group of $G_{\rm FR}$ -- denoted ${\rm Iso}(G_{\rm FR})$ -- 
is isomorphic to the the $(n+1)^2-$dimensional group $Gl(n,\mathbb{R}) \times \mathcal{H}_n$ generated by the $n^{2}$ `boosts' 
	\beq
	\label{boosts}
	p_i \hat P^j + q^j \hat Q_i-p_i q^j \xi  =p_i \frac{\partial}{\partial p_{j}} - q^j 	\frac{\partial}{\partial q^{i}} 
	\eeq
and the $2n+1$ `translations'
	\beq
	\label{translations}
	\left\{\xi,p_i \xi  -\hat Q_i,  \hat P^i-q^{i} \xi \right\}=\left\{\frac{\partial}{\partial w}, \frac{\partial}{\partial q^{i}}, \frac{\partial}{\partial p_{i}}-q^{i}\frac{\partial}{\partial w} \right\}.
	\eeq
We observe that the set \eqref{translations} satisfies the Heisenberg algebra commutation relations whose center is the Reeb vector field,
 while the `boosts' are generators of $gl(n,\mathbb{R})$  \cite{preston}.  


To define the almost para-contact structure, we notice that $\xi$ is a Killing vector field for the metric $G_{\rm FR}$ and therefore we can use the relation \cite{Zamkovoy2009}
	\beq\label{almost1}
	\nabla \xi = -\phi,
	\eeq
to find a almost contact structure which renders $G$ an associated metric. {Here, $\nabla$ is the Levi-Civita connection compatible with the metric \eqref{GinTcoordinate}.}
A direct calculation shows that \cite{TPSSASAKI}
	\beq\label{almost2}
	\phi= -\sum_{i=1}^n \left[\hat e_{(i)}^+ \otimes \hat\theta^{(i)}_{-} + \hat e_{(i)}^{-} \otimes \hat\theta^{(i)}_{+} \right]\,.
	\eeq

Therefore, the geometric structure emerging from the first two moments of the microscopic entropy change around Gibbs' distribution --  $\left(\mathcal{T},\eta,\xi,\phi,G_{\rm FR}\right)$ -- is a \emph{para-Sasakian manifold}. It can be easily shown that the curvature of the Levi-Civita connection further satisfies the $\eta$-Einstein condition, that is, its Ricci tensor is given by
	\beq
	\text{Ric} = -(2n+2) \eta \otimes \eta + 2 G_{\rm FR}.
	\eeq
Moreover, in addition to the Levi-Civita connection, one can build another connection compatible with all the defining tensors of the para-Sasakian structure. Remarkably, such a connection is flat.  Hence, it can be formally shown that the Thermodynamic Phase Space is locally isomorphic to the Hyperbolic Heisenberg Group (c.f. \cite{TPSSASAKI} and \cite{IVZ} for all the details).

\subsection{Legendre sub-manifolds: the equilibrium sub-spaces of thermodynamic systems}

The set of coordinates $\{w,q^{a},p_{a}\}$ has a natural thermodynamic interpretation on the integral sub-manifolds of the TPS. 
Of special interest are the \emph{maximal} integral sub-manifolds, $\mathcal{E} \subset \mathcal{T}$, 
i.e. those of maximal dimension which can be embedded in $\mathcal{T}$ such that their tangent bundle is completely contained in the distribution $\mathcal D$. 
These are called \emph{Legendre sub-manifolds}.

{A local description of Legendre sub-manifolds can be given as follows \cite{Arnold}.
Consider a disjoint partition $I\cup J$ of the set of indices $\{1,\dots,n\}$ and a function of $n$ variables $f(p_{i},q^{j})$,
with $i\in I$ and $j\in J$. The $n+1$ equations 
\beq\label{LegsubArnold}
q^{i}=\frac{\partial f}{\partial p_{i}} \qquad p_{j}=-\frac{\partial f}{\partial q^{j}} \qquad w=f - p_{i}\frac{\partial f}{\partial p_{i}}
\eeq
define a Legendre sub-manifold $\mathcal E$ of $(\mathcal T,\mathcal D)$.
Conversely, any Legendre sub-manifold is locally defined by these equations for at least one of the $2^{n}$ possible choices of the partition of the set $\{1,\dots,n\}$.

Notice that $f=f(p_{i},q^{j})$ in \eqref{LegsubArnold} can be any function of $n$ variables. 
In the thermodynamic interpretation, Legendre sub-manifolds represent the manifolds of equilibrium
states of a given system. Moreover, it results from \eqref{LegsubArnold} that $f(p_{i},q^{j})$ should be interpreted as the fundamental equation
for the thermodynamic potential describing a particular system 
and that \eqref{LegsubArnold} gives also the equations of state.
With this interpretation, the $2^{n}$ different possibilities  representing a given Legendre sub-manifold stand for the $2^{n}$
possible ensembles that one can, at least formally, define. 
To fix the notation, we will consider here $f$ depending only on the $q^{a}$, unless otherwise explicitly stated 
and identify $w$ with $f$ on $\mathcal E$, according to \eqref{LegsubArnold}. Therefore, a Legendre sub-manifold is defined by means of an embedding}
	\beq
	\varphi:\ \mathcal{E} \longrightarrow \mathcal{T},
	\eeq
mapping 
	\beq\label{embedding1}
	 q^a \mapsto \left[w(q^a), p_{b}(q^a), q^a\right]
	\eeq
and satisfying the isotropy condition
	\beq\label{firstlaw1}
	\varphi^* \eta = \left[\frac{\partial}{\partial q^a}w(q^b)  + p_a \right] \d q^a = 0.
	\eeq
Note that this is equivalent to demanding that the system satisfies the First-Law of thermodynamics
	\beq\label{firstlaw2}
	\d w(q^b) =- p_a \d q^a, \quad \text{where} \quad p_a = -\frac{\partial}{\partial q^a}  w(q^b).
	\eeq
Therefore, let us call  $\eta$ [c.f. equation \eqref{1stform}] the \emph{Gibbs $1$-form} and $\mathcal E$ the \emph{equilibrium manifold}.
On the equilibrium manifold the coordinate $w$ can be interpreted as a thermodynamic potential 
and the definition of the $p_a$ -- equation \eqref{firstlaw2} -- corresponds to the set of equations of state. 

Finally, the metric $G_{\rm FR}$ can be pulled back by means of \eqref{embedding1} to the Legendre sub-manifolds obtaining
\beq\label{gFR}
g=\varphi^{*} G_{\rm FR} =\frac{\partial^{2} w}{\partial q^{a} \partial q^{b}} \,\d q^{a}\otimes \d q^{b}\,.
\eeq
We observe that different choices of the embedding \eqref{embedding1} give in principle different Legendre sub-manifolds, each one equipped with
its own induced metric,
 given by the Hessian of the corresponding potential. We will analyze this aspect in more detail in \ref{LegSymm}.

\section{Contact symmetries}
\label{sec:gaugeandlegendre}
Transformations leaving the contact distribution $\mathcal D$ unchanged determine diffeomorphisms between the sub-manifolds.
Therefore, from the thermodynamic point of view, there is a large group of transformations acting on $\mathcal T$ that 
leave the equilibrium sub-manifold $\mathcal E$ unchanged, at least as long as we do not consider the induced Riemannian structure on $\mathcal E$.
In this section we study two particular examples of transformations that preserve the contact distribution, that is, gauge transformations
and Legendre symmetries.

\subsection{Gauge Transformations}\label{sec:gaugetransf}
 
We have said that the contact distribution over a contact manifold is given by $\mathcal D=\text{ker}(\eta)$ for some $\eta$ in an equivalence
class with respect to multiplication by a conformal factor. Therefore $\cal D$ is invariant with respect to a different choice of the Gibbs 1-form in the same
equivalence class.
Let us consider  transformations multiplying the 1-form $\eta$ by a conformal factor, i.e. a contact transformation for the contact structure. 
We say that a mapping $f:\mathcal T \rightarrow \mathcal T$ is a \emph{contact transformation or contactomorphism of $\eta$} if 
\beq\label{hometa}
f^{*} \eta =\Omega\,\eta
\eeq 
for some non-vanishing function $\Omega$. 
When $\Omega = 1$ we call it a \emph{strict contact transformation}. From the definition \eqref{hometa} it is clear that the contact structure $\mathcal D$ is preserved by any contact transformation. 
In particular, there are diffeomorphisms that leave the contact structure $\mathcal D$ invariant. We say that such a diffeomorphism 
$f:\mathcal T\rightarrow \mathcal T$ is an \emph{infinitesimal contact transformation} if 
\beq\label{contactomorphisms}
\pounds_{X_{f}}\eta=\tau\,\eta\,,
\eeq
where $\tau:\mathcal T\rightarrow \mathbb R$ is a non-vanishing function and $\pounds_{X_{f}}\eta$ is the Lie derivative of $\eta$ along 
the flow generated by the infinitesimal displacements corresponding to $f$.
Then an infinitesimal contactomorphism is strict if and only if $\pounds_{X_{f}}\eta=0$, that is, $f$ not only leaves the contact structure invariant, but also its representative 1-form. In this sense we say  
that a strict contactomorphism is a \emph{symmetry of the contact form}, while a general contactomorphism can be regarded as a conformal symmetry. 
{Finally, notice that strict contactomorphisms are also called \emph{quantomorphisms} in some recent literature on contact Riemannian geometry, c.f. \cite{edin1}.}

Given $\mathcal D$, we consider now the splitting \eqref{splitting}. 
This splitting is not unique, as $\xi$ depends on the particular choice of $\eta$ [c.f. equation \eqref{Reeb}].
In particular, the first condition in \eqref{Reeb} just implies that  $\xi$ needs to be re-scaled when one changes $\eta$. 
More complicated is the change due to the second condition in \eqref{Reeb}, which can also change the direction of $\xi$, as we will shortly see. Therefore
the splitting \eqref{splitting} changes in a non-trivial manner.
Notice that if the 1-form $\eta$ is transformed by means of a strict contactomorphism, then the splitting of the tangent bundle remains unchanged. 
That is the reason why we say that strict contactomorphisms are symmetries of the contact bundle.

Let $(\mathcal T,\eta,\xi, \phi,G)$ be a para-contact metric manifold and choose a different $1$-form $\tilde\eta$ in the same equivalence class of $\eta$.
Note that it must be $\tilde\eta = \Omega\,\eta$ where $ \Omega$ is an everywhere non-vanishing function on $\mathcal T$. 
Obviously the contact distribution of the two is the same [c.f. equation
\eqref{xi}]. However the Reeb vector field, the almost para-contact and the metric structure \emph{depend} on the choice of $\eta$.
{In fact, it turns out that the para-contact metric  structure associated to $\tilde\eta$ is obtained by \cite{Zamkovoy2009}
	\begin{align}
	\label{gaugetransf1}
	\tilde\xi  & 		=\frac{1}{\Omega}\left(\xi+\zeta\right),\\
	\label{gaugetransf2}	
	\tilde\phi &	= \phi +\frac{1}{2\Omega}\,\eta\otimes\left[G^{-1} (\d\Omega, \cdot ) - \xi (\Omega)\, \xi\right],\\
	\label{gaugetransf3}	
	\tilde G 	&		= \Omega \left( G-\eta\,{\otimes} \,z-z\,{\otimes}\, \eta\right)+\Omega \left(\Omega-1+|\zeta|^{2}\right)\eta\otimes\eta\,,
	\end{align}
}where
	\beq
	\zeta=-\frac{1}{2\Omega}\phi\left[ G^{-1}(\d\Omega, \cdot)\right] \quad \text{and} \quad z=G(\zeta, \cdot).
	\eeq
The change from $(\mathcal T,\eta,\xi,\phi,G)$ to $({\mathcal T},\tilde\eta,\tilde\xi,\tilde\phi,\tilde G)$ is called a \emph{gauge transformation} of the para-contact metric structure.
When $\Omega$ is constant it is called a \emph{D-homothetic deformation}.

Note that if the initial manifold $(\mathcal T,\eta,\xi,\phi,G)$ is a para-contact metric manifold (resp. a para-Sasakian manifold),
 then the new structure defined as $({\mathcal T},\tilde\eta,\tilde\xi,\tilde\phi,\tilde G)$ is still a para-contact metric manifold  (resp. a para-Sasakian manifold). 
However, as we see from equations \eqref{gaugetransf1}-\eqref{gaugetransf3}, even though the contact 1-form scales just by a function,  the associated Reeb vector field,  almost para-contact
structure and metric tensor all change by non-trivial transformations. 


Let us see an example which is relevant in ordinary thermodynamics. 
Consider the Gibbs 1-form generating the First Law of thermodynamics in
the molar internal energy representation together with its associated metric, that is
\beq\label{1stformex}
\eta^{U}=\d U-T\,\d S+p\, \d V\,.
\eeq
and
\beq\label{GRexample}
G^U_{\rm FR} = \eta^U \otimes \eta^U + \d T \overset{\rm s}{\otimes} \d S - \d p \overset{\rm s}{\otimes} \d V.
\eeq
{Notice that comparing equations \eqref{1stform} and \eqref{1stformex}, the coordinates $\{w,q^{1},q^{2},p_{1},p_{2}\}$ here are given by $\{U,S,V,-T,p\}$.}
It is well-known that we can change to the entropy representation just by multiplying $\eta^{U}$ by a conformal factor $\Omega=-1/T$. 
Thus one obtains another 1-form in the same equivalence class which reads
\beq
\eta^{S}= -\frac{1}{T} \eta^U= \d S-\frac{1}{T}\,\d U-\frac{p}{T}\, \d V\,.
\eeq 
{Moreover, in this case the almost para-contact structure \eqref{almost2} associated to $G^{U}_{\rm FR}$ according to \eqref{associatedG} reads}
\begin{equation}
\phi^{U}
=-\left(T\,\d S-p\,\d V\right)\otimes \frac{\partial}{\partial U}-\d S\otimes \frac{\partial}{\partial S}-\d V\otimes \frac{\partial}{\partial V}+\d T\otimes \frac{\partial}{\partial T}+\d p\otimes \frac{\partial}{\partial p}.
\end{equation}
Using $\phi^{U}$ and $G^{U}_{\rm FR}$ we can compute explicitly the gauge transformation \eqref{gaugetransf1}-\eqref{gaugetransf3} with $\Omega=-1/T$ to obtain
the change in the para-Sasakian structure associated to the change of representation 
\begin{align}
	\label{gaugetransf1ex}
	\xi^{S}  & 	=\frac{\partial}{\partial S}\\
	\label{gaugetransf2ex}	
	\phi^{S} &	= \phi^{U} -\frac{1}{T}\,\eta^{U}\otimes\left[T\frac{\partial}{\partial U}+\frac{\partial}{\partial S}\right]=\phi^{U} -\frac{1}{T}\,\eta^{U}\otimes\hat{Q}^{1}\nonumber\\
	& =-\left(\frac{1}{T}\d U+\frac{p}{T}\d V\right) \otimes\frac{\partial}{\partial S}-\d U\otimes \frac{\partial}{\partial U}-\d V\otimes \frac{\partial}{\partial V}+\d T\otimes \frac{\partial}{\partial T}+\d p\otimes \frac{\partial}{\partial p},\\
	\label{gaugetransf3ex}	
	 G_{\rm FR}^{S} & = -\frac{1}{T} \left( G_{\rm FR}^{U}+\frac{1}{T}\,\eta^{U}\overset{s}{\otimes} \d T\right)+\frac{1}{T} \left(\frac{1}{T}+1\right)\eta^{U}\otimes\eta^{U}\nonumber\\
	 & =\eta^{S}\otimes\eta^{S}+\d U\,\overset{s}{\otimes} \,\d\left(\frac{1}{T}\right)+\d V\,\overset{s}{\otimes} \,\d\left(\frac{p}{T}\right).
\end{align}
}

Accordingly, one obtains two different metric structures on the Legendre sub-manifold $\mathcal E$.
Let us call such metrics $g^{U}=\varphi_{U}^{*}\, G_{\rm FR}^{U}$ and $g^{S}=\varphi_{S}^{*}\, G_{\rm FR}^{S}$, respectively. 
Therefore, it is immediate to realize from \eqref{gaugetransf3ex} and \eqref{gFR} that  
the change in the metric structure from $G_{\rm FR}^{U}$ to $G_{\rm FR}^{S}$ due to the gauge transformation induces a conformal change from the metric
$g^{U}$ to $g^{S}$ given by
	\beq\label{fromWtoR}
	g^{S}=-\frac{1}{T}\,g^{U}\,.
	\eeq
Such change
is the well-known conformal relation between Weinhold's `energy' metric  and (minus) Ruppeiner's `entropy' metric on the equilibrium manifold 
which was first derived  in \cite{SalamonRW}. A more complete study of gauge transformations in thermodynamics will be examined in another work
\cite{usinpreparation}. 
In the next sub-section we will consider Legendre transformations and show that they indeed represent a symmetry of the contact bundle of the TPS, in the sense that they leave 
such structure unchanged.

\subsection{Legendre Symmetry}\label{LegSymm}
Let us consider now transformations leaving the Gibbs 1-form $\eta$ invariant, i.e. symmetries of the contact 1-form. 
We say that a mapping $f:\mathcal T \rightarrow \mathcal T$ is a \emph{symmetry of $\eta$} if 
\beq\label{symmeta}
f^{*} \eta = \eta\,.
\eeq 
As we have discussed in the preceding sub-section, strict contactomorphisms leave the 1-form
$\eta$ invariant, therefore they are symmetries of $\eta$. However this class is in principle larger, because
it includes also transformations not generated by an infinitesimal group of transformations. 

As we have already commented, if $\eta$ is invariant, then
 the splitting of the contact bundle \eqref{splitting} is unchanged, as well as 
 the corresponding Reeb vector field $\xi$, the almost para-contact structure
$\phi$ and the associated metric $G_{\rm FR}$. 
However, this does not mean that a symmetry of the contact 1-form is also a symmetry of the 
 metric structure, as we discuss below.

As a particular class of symmetries of the 1-form $\eta$ relevant in thermodynamics, let us consider Legendre transformations.
 A \emph{(discrete) Legendre transformation} on the TPS on $\mathcal T$ is
 given by the relations 
	\begin{empheq}[left=\empheqlbrace]{align}
	\label{LT.1}
	&\tilde \w_{(i)} \equiv  \w - q^{(i)} p_{(i)}\\ 
	&\tilde p_{(i)} 	\,\equiv \, q^{(i)} \quad  \text{and}\label{LT.2}\\
	\label{LT.3}	
	&\tilde q^{(i)}	\,\equiv  -p_{(i)},
	\end{empheq}
for $i\in I\subseteq \{1,\dots,n\}$ while leaving the rest of the coordinates unchanged, i.e. $\tilde q^j = q^j$ and $\tilde p_j= p_j$ for $j\neq i$. 
A \emph{partial Legendre transformation} (PLT) only interchanges the pairs of thermodynamic variables in the subset $I$. 
A transformation that exchanges every pair of coordinates is called a \emph{total Legendre transformation} (TLT).

Note that, as well as a change of representation is a (discrete) example of a gauge transformation, a
 Legendre transformation $f$ is a (discrete) example of a symmetry of the 1-form $\eta$. In fact, it is easy to check that $f^*\, \eta = \eta$. 
As such, it follows that the contact 1-form $\eta$ and its Reeb vector field $\xi$ are invariant, and hence the splitting of the tangent bundle \eqref{splitting}
is unchanged. We argue here that the basic equilibrium thermodynamics is completely determined by such splitting. Therefore, 
a Legendre transformation is a symmetry of equilibrium thermodynamics, as expected.

It is easy to verify that a Legendre transformation is not a symmetry of the metric structure \eqref{GinTcoordinate}.
For example, a partial Legendre transformation $f_{1}$ -- exchanging only the first pair of variables $q^{1}$ with $p_{1}$ -- changes  the metric \eqref{GinTcoordinate}  to
\beq\label{FisherRaorhoG3}
\tilde{G}_{\rm FR} = f_1^{*} G_{\rm FR} = \left(\d \tilde \w+\tilde p_a\,\d \tilde q^a\right)\otimes\left(\d\tilde\w+\tilde p_a\,\d \tilde q^a\right)
 + \d\tilde q^{1}\overset{\rm s}{\otimes}\d\tilde  p_{1}
-\sum_{i\neq1}\d\tilde  q^i\overset{\rm s}{\otimes}\d\tilde  p_i\,,
\eeq
where the second term on the right hand side has changed.
Physically, one can interpret the metric induced by  \eqref{GinTcoordinate} onto any equilibrium sub-manifold $\mathcal E$
as giving a measure of the probability of fluctuations of the unconstrained thermodynamic variables of the corresponding ensemble \cite{rupp1995}.
Moreover, a Legendre transformation represents the changing from a thermodynamic ensemble to the other, thus changing the constrained variables and 
the fluctuating ones.
Therefore, a change of the metric structure is completely equivalent to the fact that fluctuating variables and the value of the fluctuations are different in the various ensembles \cite{Callen}.
Within this interpretation, the fact that the First Law of thermodynamics -- represented by the {vanishing of the} 1-form $\eta$ -- is invariant under a Legendre transformation proves
that the mean values of the thermodynamic functions do not change with the election of ensemble. However, the situation is not the same for the values of the fluctuations of such 
functions, and thus for the corresponding metric structure in the geometric construction.
In some formulations of the geometry of thermodynamics it has been further required that the metric structure of the TPS should be \emph{invariant} with respect
to Legendre transformations (see e.g. \cite{GTD, CRGTD}). However, we will not consider such requirement here, as we are working  
with the metric structure derived from Gibbs' statistical mechanics and information theory as in \cite{MNSS1990,TPSSASAKI} and corresponding to thermodynamic fluctuation theory \cite{rupp1995}.
However, it would be worth to explore if different forms of the entropy functional in statistical mechanics (e.g. R\'enyi or Tsallis entropies \cite{renyi,Tsallis2}) can lead to other types of thermodynamic
metrics, in the same way as one derives the metric \eqref{GinTcoordinate} directly from the Boltzmann-Gibbs entropy functional \cite{TPSSASAKI}.

In the next section we will see the physical implications of the Legendre symmetry of the TPS in the geometric description of ordinary equilibrium thermodynamics.
 
\section{Legendre symmetry and equivalence of the ensembles}
\label{sec:Lsym}

As we have seen in the previous section, a Legendre symmetry preserves the splitting of the tangent bundle of the TPS. 
Therefore, it induces a diffeomorphism between its Legendre sub-manifolds, as we show here. 
This is the formal cause of the well-known fact that we can use the equations of state to perform a Legendre transformation, 
changing the thermodynamic potential and exchanging the role of  the independent variables of the system. 
Here we show that this is always possible as long as the potential satisfies the global convexity conditions. 
In ordinary homogeneous thermodynamics this requirement is equivalent to say that the system is in a single phase \cite{Callen}. 

\subsection{Legendre symmetry as a diffeomorphism}\label{subsec:LegSym}

{Let us now consider the embedding \eqref{embedding1}
with two different choices of the thermodynamic potential.
For simplicity, let us consider $w(q^{a})$ and $\tilde\w(p_{a})$, with $\tilde\w(p_{a})$ the total Legendre transform of $\w$ [c.f. equation \eqref{LT.1}].} 
These different choices in principle induce two different Legendre sub-manifolds $\mathcal E^{\w}$ and $\mathcal E^{\tilde\w}$, respectively.  
We show here that the Legendre symmetry of the TPS induces a diffeomorphism 
	\beq\label{Ldiffeo}
	\psi\,\,:\,\, \mathcal E^{\w}\longrightarrow\,\,\mathcal E^{\tilde\w}
	\eeq
if and only if 
	\beq\label{Hneq0}
	\frac{\partial^{2}\w}{\partial q^a\partial q^b}\neq0 
	\eeq 
at every point.
Such transformation induces -- by means of equations  \eqref{embedding1} and \eqref{firstlaw1} -- 
a diffeomorphism 
	\beq\label{Ldiffeo2}
	\begin{split}
	\psi\,\,\,:\,\,\mathcal E^{\w} \,\,\, & \longrightarrow \,\,\, \mathcal E^{\tilde\w}\\
	\left[\w(q^a),q^a,p_a(q^b)\right] &\,\, \mapsto \left[\tilde\w(p_a),q^a(p_b),p_a\right]
	\end{split}
	\eeq 
that
 transforms the thermodynamic potential from $\w(q^a)$ to $\tilde \w(p_a)$
and at the same time
 interchanges the role of the independent coordinates from $q^a$ to $p_a$. 
 The explicit expression of the transformation $\psi$ is given by the equations of state 
	\begin{equation}\label{Eos}
	p_a(q^b)=-\frac{\partial}{\partial q^a}\w(q^b).
	\end{equation}
It is straightforward  to calculate the push-forward of such transformation, which is
	\begin{equation}\label{tangentmap}
	\begin{split}
	&\psi_*\,:\,\,\,T\mathcal E^{\w}\,\,\longrightarrow \,\,\,T\mathcal E^{\tilde\w}\\
	&\psi_* X \equiv -\left(\frac{\partial^{2}\w}{\partial q^a\partial q^b}\right)\,X^b\frac{\partial}{\partial {p_{a}}}\,,
	\end{split}
	\end{equation}
where $X=X^{a}\partial_{q^a}$ is any vector field on $T\mathcal E^{\w}$. 
Therefore, we see from equation \eqref{tangentmap}
that, although a Legendre transformation  is always a symmetry of the TPS, it induces  
a diffeomorphism on the equilibrium sub-manifolds {\it if and only if the Hessian of the potential $\w(q^a)$ is non-degenerate}. 
Therefore, such diffeomorphism between the equilibrium sub-manifolds depends on the particular function $\w(q^a)$, i.e.
depends on the particular system under exam.
Whenever such Legendre symmetry of the TPS is broken on $\mathcal E$, then the transformation on $\mathcal E$
corresponding to a Legendre transformation is not a diffeomorphism. In particular, in such case the sub-manifolds $\mathcal E^{\w}$ and 
$\mathcal E^{\tilde\w}$ are not equivalent.
  The in-equivalence of the information contained in $\mathcal E^{\w}$ and $\mathcal E^{\tilde\w}$ in such case 
reflects geometrically the in-equivalence of the two ensembles,
which is well-known in statistical mechanics and thermodynamics for regions of the phase diagram where the thermodynamic potential does not 
satisfy  the \emph{global} convexity conditions, that is, where it is not a concave [resp. convex] function of the extensive variables \cite{Callen}.
{Notice that in this case the whole structures of the Legendre sub-manifolds are inequivalent, not just their metrics.}

Consider for example the ideal gas. 
This is a system whose thermodynamic potential $\w(q^a)$ globally satisfies the concavity conditions and therefore 
the ensembles are completely equivalent over the full region of the phase diagram. 
This is represented in contact geometry by a single, smooth, Legendre sub-manifold $\mathcal E$ (up to diffeomorphisms).
However, the majority of systems undergo instabilities and have regions where different phases coexist and hence we have different equilibrium
sub-manifolds $\mathcal E$ corresponding to the different ensembles, and these can intersect over the regions of coexistence. 
In ordinary thermodynamics, one recovers the global stability (the concavity requirement) by means of the Maxwell equal area law, but the (local) 
Legendre symmetry cannot be restored \cite{Callen}.  

As we have seen, the breaking of the Legendre symmetry allows for the existence of different ensembles, represented by different sub-manifolds $\mathcal E$
whose information over the region of coexistence is inequivalent.
Besides the in-equivalence of the ensembles, 
the intersection of such sub-manifolds implies also that there are processes that can pertain to different phases at the same time.
As we will shortly see,  such processes represent physically that the system is going from a thermodynamic phase into a different one, following a sequence of equilibrium states, 
i.e. the system is undergoing a \emph{coexistence process}. 
In the sub-section \ref{firstorderphtrans} 
we give a geometric characterization of such processes as curves on the $N$-dimensional sub-manifolds lying in the intersection 
of $r$ (equilibrium) Legendre sub-manifolds denoting the $r$ different coexisting phases. 
The dimension $N$ is calculated by Gibbs'
phase rule and it turns out that 
\beq
N=C-r+2,
\eeq
 where $C$ is the number
of different species in the thermodynamic system. 
In particular, for most of the cases in classical thermodynamics of simple systems (i.e. $C=1$), 
the coexisting region is $1$-dimensional, i.e. a curve, in the case of two coexisting phases ($r=2$)
and $0$-dimensional, i.e. a point, in the case of three coexisting phases ($r=3$), as e.g. in the triple point of water \cite{Callen}.

There is a subtle point to be highlighted here, regarding the non-degeneracy of the Hessian of the potential, equation \eqref{Hneq0}. 
Whenever we consider an ordinary thermodynamic system for which the entropy (or the internal energy) is a homogeneous function of order one
of the extensive variables,
then, due to the Gibbs-Duhem relation \eqref{w1law},
the total Legendre transformation is always degenerate, an indication that we are considering more degrees of freedom than necessary. 
Thus, one uses the scaling property of the system to fix one of the extensive variables and divides the rest of them by such a fixed amount. 
The result of this operation also divides the potential by the same amount. 
In practice, one either chooses the particle number or the volume, and works with molar quantities or densities, respectively, for which the Legendre transformation is well-defined.

\subsection{Legendre symmetry as an isometry}
Consider now the metric structure induced on $\mathcal E$. We have said that different choices of the thermodynamic potential
$w(q^{a})$ in \eqref{embedding1}
can induce on $\mathcal E$ different and in principle in-equivalent metric structures defined as the Hessian of the corresponding potential.
Here we show that the \emph{total} Legendre transform always induces an isometry between the corresponding induced metric structures. 
{This result is known in Hessian geometry (c.f. \cite{HessianGeomBook} p. 27), we re-derive it here in the context of thermodynamics to make direct contact with our discussion of the Legendre
transform as a symmetry of the geometric structure of thermodynamic fluctuation theory. Moreover, we point out that the same is not
true 
for a \emph{partial} Legendre transform.
}

Consider for example the metrics induced on $\mathcal E$ by the embedding \eqref{embedding1} and corresponding to $w$ and its total Legendre transformation
$\tilde w$, as defined in \eqref{LT.1}. For clarity, we re-write explicitly the corresponding embeddings
\beq\label{embedw}
\varphi_{w}:(q^{a})\mapsto\left[w(q^{a}),q^{a},p_{a}\right], \qquad p_{a}=-\frac{\partial}{\partial q^{a}}w
\eeq 
and 
\beq\label{embedwtilde}
\varphi_{\tilde w}:(p_{a})\mapsto\left[\tilde w(p_{a}),q^{a},p_{a}\right], \qquad q^{a}=\frac{\partial}{\partial p_{a}}\tilde w.
\eeq
The two embeddings \eqref{embedw} and \eqref{embedwtilde} define the two metric structures 
\beq\label{gw}
g=\varphi_{w}^{*} G_{\rm FR} = -\frac{\partial^{2} {w}}{\partial q^{a} \partial q^{b}}\, \d q^{a} \otimes \d q^{b}
\eeq 
and 
\beq\label{gwtilde}
\tilde g=\varphi_{\tilde w}^{*} G_{\rm FR} = \frac{\partial^{2} {\tilde w}}{\partial p_{a} \partial p_{b}} \,\d p_{a} \otimes \d p_{b}.
\eeq

These two metrics are in principle different. Let us see that a total Legendre transformation on $\mathcal T$ induces an isometry between the two.
In general, we say that a diffeomorphism $\psi:(\mathcal E^{w},g)\rightarrow (\mathcal E^{\tilde w},\tilde g)$ is an \emph{isometry} if 
	\beq\label{isometrydef1}
	\psi^* \tilde g = g.
	\eeq
In general we have that under a diffeomorphism $\psi$ the action of an induced pullback $\psi^*:T^*\mathcal E^{\tilde w} \to T^*\mathcal{E}^w$ on the metric is defined by
	\beq\label{isometrydef2}
	\psi^* \tilde{g}(X,Y) =\tilde{g}(\psi_{*}X,\psi_{*}Y) \quad \text{for any} \quad X,Y \in T\mathcal E,
	\eeq   
where $\psi_*: T\mathcal{E}^w \to T\mathcal{E}^{\tilde w}$ is the differential map. Consider now the diffeomorphism on $\mathcal{E}$ induced by a total Legendre transformation as defined in \eqref{Ldiffeo2}. Calculating the components for $\tilde g$ with respect to pushforward of the basis in $T\mathcal{E}$ we obtain
	\begin{align} \label{gtildecomp}
	\tilde g\bigg(\psi_* \frac{\partial}{\partial q^a}, \psi_* \frac{\partial}{\partial q^b}\bigg) &= 
	\frac{\partial^2 \tilde w}{\partial p_c \partial p_d} \d p_c \otimes \d p_d 		\bigg(\frac{\partial p_e}{\partial q^a}\partial_{p_e}, \frac{\partial p_f}{\partial q^b} \partial_{p_f}\bigg) \nonumber\\ 
	&=  
	\frac{\partial^2 \tilde{w}}{\partial p_c \partial p_d} \frac{\partial p_c}{\partial q^a}\frac{\partial p_d}{\partial q^b} = \bigg(\frac{\partial q^d}{\partial p_c}\bigg) \frac{\partial p_c}{\partial q^a}\frac{\partial p_d}{\partial q^b}
	\nonumber  \\ 
	&= \frac{\partial p_a}{\partial q^b} = -\frac{\partial^2 w}{\partial q^a \partial q^b},
	\end{align}
where we have used equations \eqref{tangentmap}, \eqref{embedw} and  \eqref{embedwtilde}. 
We observe that the components of $\psi^{*} \tilde g$ and $g$ are the same with respect to the basis of $T \mathcal{E}$, 
thence we have proved that the diffeomorphism induced by a total Legendre transformation on the equilibrium manifolds is also an isometry. 
However, the same is not true if we consider a partial Legendre transformation. 
This can be seen by the fact that the scalar curvatures of the two structures are in general different (see e.g. 
\cite{iofra} for their comparison).

To conclude, let us remark that we have considered here each Legendre sub-manifold $\mathcal E$ as equipped with the natural induced metric from the metric $G_{\rm FR}$ on the TPS 
and we have analyzed the conditions for these metrics to be equivalent.

\subsection{First order phase transitions}\label{firstorderphtrans}

\begin{figure}[h!]
	\begin{center}
	\includegraphics[width=0.45\columnwidth]{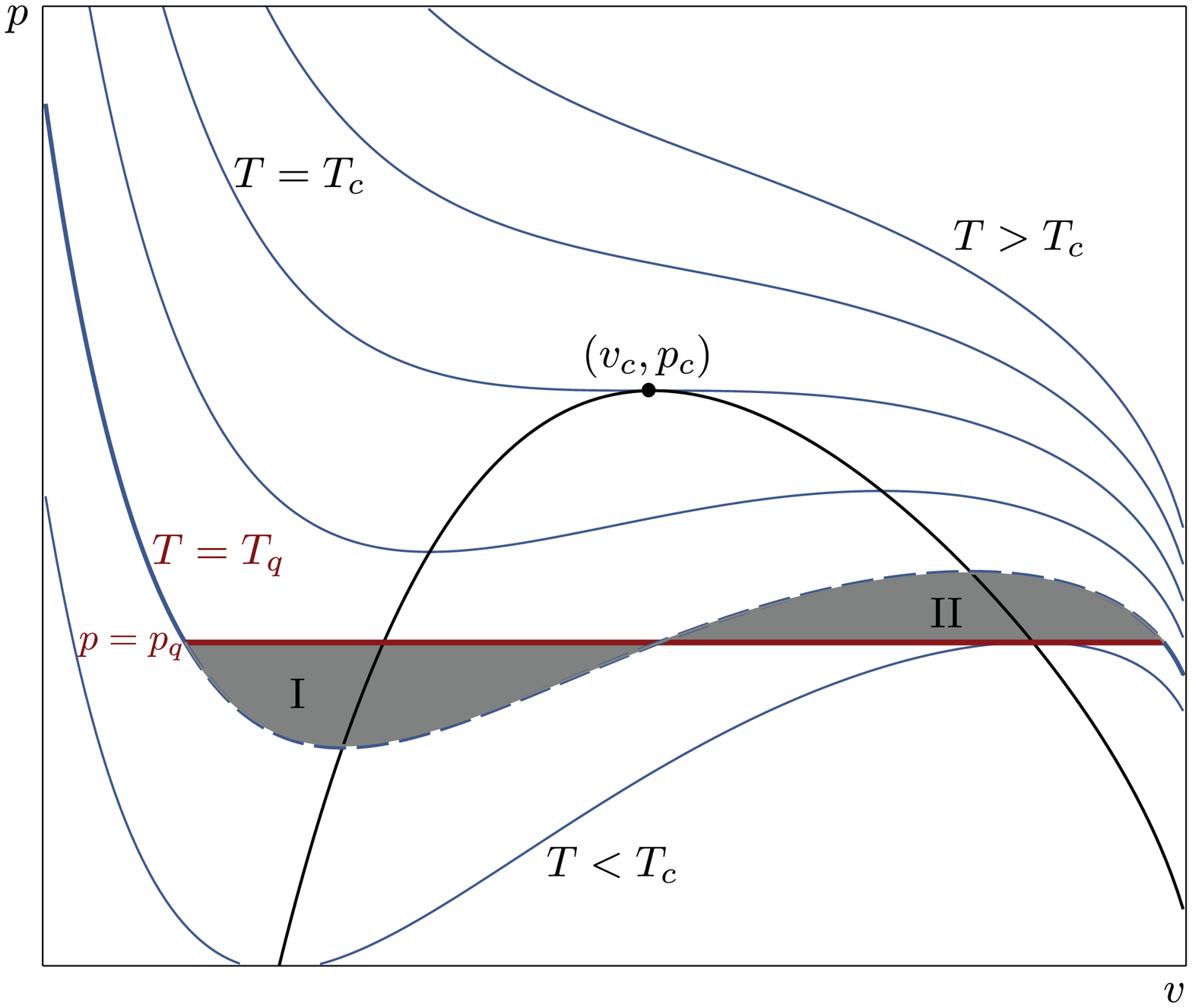}\hskip1cm\includegraphics[width=0.45\columnwidth]{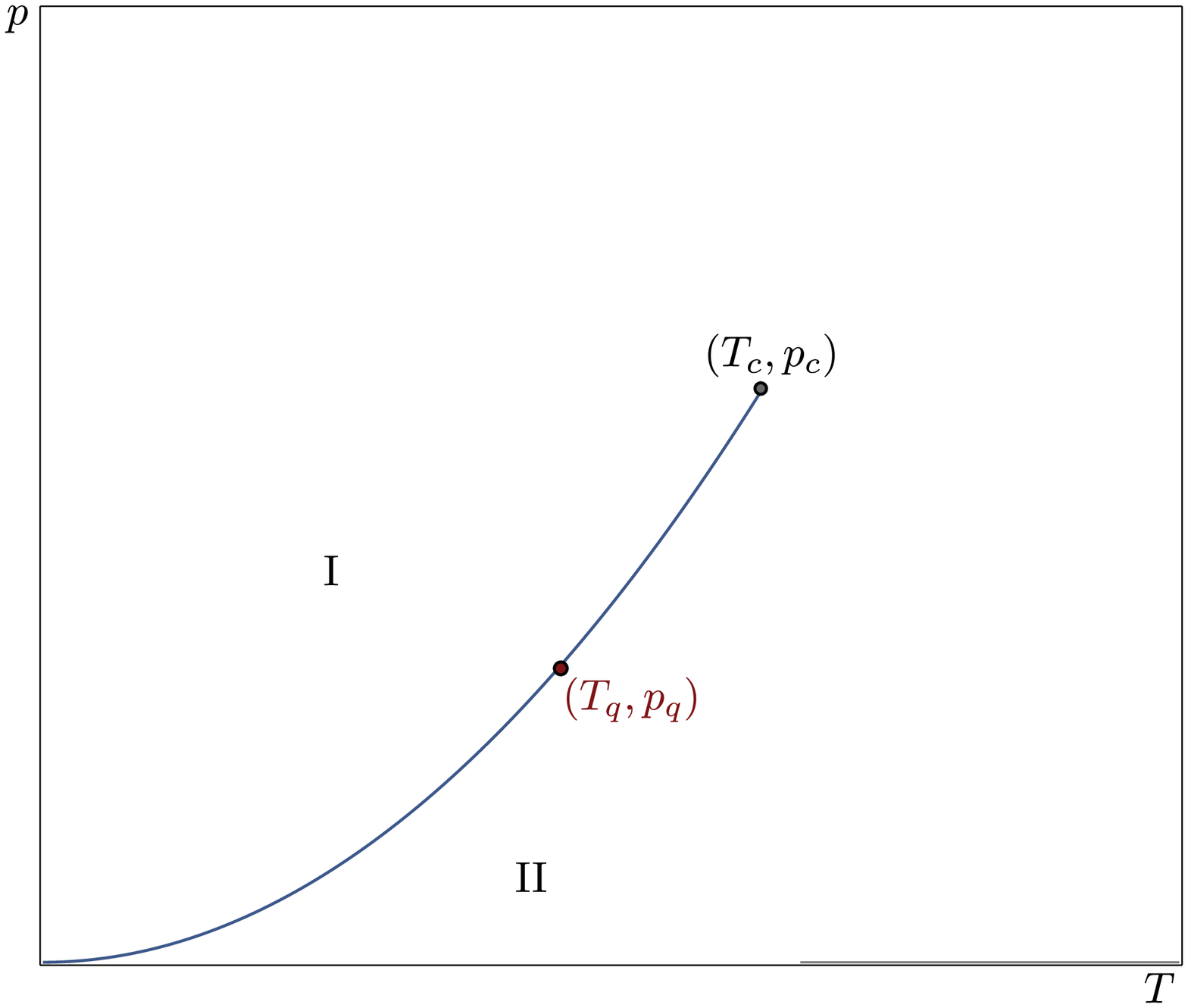}
	\end{center}
	\caption{The coexistence process as it appears on a $P-V$ diagram and on a $P-T$ diagram. Two aspects are of major interest. The first is to note that
	the process of coexistence (in red in the two panels) is represented by a line in the $P-V$ diagram and by a single point in the $P-T$ diagram. Second, in order for the two phases to coexist 
	at equilibrium, the temperatures and the pressures must be the same. More details in the text. {Notice that we have also highlighted the \emph{spinodal}
	curve (it is depicted in black in the $P-V$ diagram), indicating the points where the local stability conditions fail at any given temperature. Finally, the critical point $(v_{c},p_{c},T_{c})$ 
	is depicted in both panels.}}
	\label{figure1}
	\end{figure}

Let us now turn to describe {another aspect that has received little attention in the geometric descriptions of thermodynamics. We investigate here
the only region of the phase space} where one encounters ensemble in-equivalence in ordinary thermodynamics, {namely} the region of coexistence
between different phases (for a more detailed description see e.g. \cite{Callen}). 
To this end, we refer to the {Pressure-Volume and Pressure-Temperature} diagrams of the liquid-vapour coexistence for a Van der Waals fluid presented in Fig.~\ref{figure1}. 
Above the critical temperature $T_c$, the isotherms in the $P-V$ diagram are decreasing functions of $V$ and therefore are stable. 
On the contrary, below the critical temperature, the isotherms have a region of instability, which is ``cut-out'' by means of the Maxwell construction (the red horizontal line in the left panel of Fig.~\ref{figure1}). 
Such construction consists in finding the equilibrium value for the pressure at which the two phases coexist at equilibrium. 
It turns out that such equilibrium value is given by requiring that the two areas indicated by I and II in the $P-V$ diagram be equal. 

It is worth noting that, when a coexistence of two or more phases is present, already at
the level of ordinary thermodynamics, we can see that the description in the variables $q^a$ is {\it not equivalent} to that using the variables $p_a$. 
In fact, by looking at the two diagrams in Fig.~\ref{figure1}, one immediately sees that in the $P-V$ coordinates the coexistence process is given by a straight line (in red), 
whereas in the $P-T$ coordinates it corresponds to the single red point indicated by $(T_{q},p_{q})$. Consequently, it is not surprising that 
the change of coordinates by a Legendre transform in that region is not a diffeomorphism. 
Indeed, as we have already pointed out, the descriptions using the extensive or the intensive coordinates are equivalent  
\emph{if and only if the Legendre transform is well defined}, i.e. when the global concavity conditions are strictly satisfied \cite{Callen}.

From the above observations, we derive the conclusion that
the process of coexistence cannot be fully described on a specific equilibrium manifold $\mathcal E$. This 
is because the region of coexistence is the intersection of different equilibrium sub-manifolds and the Legendre transformation is not a diffeomorphism along such intersection.
However, we can solve this problem in a simple way.
{It is usually assumed that equilibrium states are the ones
belonging to a particular Legendre sub-manifold.}
However, more generally we can {characterize} any {equilibrium}  process as a  parametrized curve $\gamma(t)$ on the TPS  {satisfying the First Law at every point along the path}, that is 
	\beq
	\label{newgammadot}
	\eta(\dot \gamma)  = 0,
	\eeq
{where $\dot \gamma$ denotes the tangent vector to the curve at the point of evaluation.}
In particular we can do so for coexistence processes.
Therefore from now on we will look at all equilibrium processes as curves on the TPS whose tangent vector annihilates the 1-form $\eta$. 
By this definition, processes of coexistence of different phases are normal equilibrium processes and can be described in the geometric framework.

In the next section, after a short review of contact Hamiltonian geometry, we will introduce
  a \emph{contact} Hamiltonian formulation of thermodynamics which parallels the \emph{symplectic} Hamiltonian formulation of conservative
mechanics. In particular, we will define a contact Hamiltonian function that is the analogue of the Hamiltonian energy
for mechanical systems. In fact, the flow of such function defines the evolution of the thermodynamic system, i.e. thermodynamic processes.
Remarkably, we will see that the relevant contact Hamiltonian in thermodynamics coincides -- up to a sign -- with the entropy of the system, considered as
a function on the TPS, equation \eqref{macroEnt}. 
Contrary to symplectic mechanics, we will see that in this case the Hamiltonian is conserved only on a particular sub-class of processes, i.e. equilibrium processes.

\section{Contact Hamiltionian Thermodynamics}\label{sec:ECH}

In this section, after a brief review of the main aspects of contact Hamiltonian dynamics, we use
these concepts to
introduce a Hamiltonian function on the TPS whose flow defines thermodynamic processes. 
{Such a formulation was intended first by Mrugala (c.f. for instance \cite{mrugala2000,mrugalachina,mrugalaJPA}), there he presented some special cases,
valid only for particular thermodynamic systems. Moreover, Rajeev in \cite{rajeevHJ} has also given a contact Hamiltonian formulation of thermodynamics, 
based on the Hamilton-Jacobi formalism. However,  the characteristic curves of the generating functions that he considers
 give the equations of state of the substance and therefore, although the construction is very neat, there is no real `time evolution' of the system. 
Here we want to propose an approach similar to standard Hamiltonian mechanics, and therefore we demand the flow of the Hamiltonian function
 to define thermodynamic processes.}


\subsection{Brief review of contact Hamiltonian dynamics}

To begin with, let us briefly review  some contact Hamiltonian dynamics, following in particular \cite{boyerintegrablesystems}.
{We start with the contact manifold $\mathcal T$ in which the representative contact 1-form $\eta$ is fixed. Therefore, to every}
differentiable function $h:\mathcal{T} \rightarrow \mathbb{R}$, we can associate a vector field $X_{h}$, called the \emph{Hamiltonian vector field generated by $h$},
defined {through Cartan's identity 
	\beq\label{cartan}
	\pounds_{X_{h}} \eta = \d \eta \left(X_{h},\cdot\right) + \d \left[\eta(X_{h}) \right](\cdot),
	\eeq
and the relation
	\beq\label{isomorphismLiealg}
	h = \eta\left(X_h\right).
	\eeq
In local Darboux coordinates, the Hamiltonian vector field $X_{h}$ is given by
	\beq
	\label{generic.ham}
	X_h = \left( h - p_a \frac{\partial h}{\partial p_a}\right)\frac{\partial}{\partial \w} + \left(
	p_a \frac{\partial h}{\partial \w}-\frac{\partial h}{\partial q^a} \right)\frac{\partial }{\partial p_a} + \left(\frac{\partial h }{\partial p_a} \right)\frac{\partial }{\partial q^a}.
	\eeq	
{while in the {Heisenberg} basis \eqref{basisTT} it takes the much simpler form
	\beq\label{Xhcanonical}
	X_{h}=h\, \xi +\hat{Q}_{i}(h)\hat{P}^{i}-\hat{P}^{i}(h)\hat{Q}_{i}\,.
	\eeq 
{It turns out that
	\beq\label{Leta}
	\pounds_{X_{h}}\eta=\xi(h)\,\eta
	\eeq
and thus the Cartan identity \eqref{cartan} can be written {using \eqref{isomorphismLiealg} and \eqref{Leta}} as
	\beq
	\label{defham}
	\d \eta \left(X_{h},\cdot\right) = - \d h + \xi(h)\, \eta.
	\eeq
Notice the contrast with the general criterion to define  Hamiltonian systems over symplectic manifolds, 
where the symplectic two-form operating on the vector field generated by the Hamiltonian function must be an exact differential whereas, in the contact case, the left hand side of \eqref{defham} is not necesarily an exact form. Thus, we say $h$ is a \emph{contact Hamiltonian}. 

Hamiltonian vector fields form exactly the Lie algebra ${\bf \it con}(\mathcal T, \mathcal D)$ of contactomorphisms, c.f. equation \eqref{contactomorphisms}.
When $h$ is a \emph{basic function}, i.e. $\xi(h)=0$ with $\xi$ the Reeb vector field, they reduce to the sub-algebra  ${\bf \it con}(\mathcal T, \mathcal \eta)$  
of strict contactomorphisms, or symmetries of $\eta$.  
 The mapping \eqref{isomorphismLiealg}, sending every vector field $X\in {\bf \it con}(\mathcal T, \mathcal D)$ to the corresponding function
 $\eta(X)\in C^{\infty}(\mathcal T)$  is an isomorphism of Lie algebras, where the Lie algebra structure of $C^{\infty}(\mathcal T)$ is given by the \emph{Jacobi
 bracket}
 	\beq\label{Jacobibr}
	\left\{\eta(X),\eta(Y)\right\}_{\eta}=\eta\left([X,Y]\right).
	\eeq   

Note that both the isomorphism and the definition of the Jacobi bracket depend crucially on the choice of the Gibbs 1-form $\eta$.
Notice as well that the Reeb vector field \eqref{Reeb} associated with the contact form $\eta$  is the Hamiltonian
vector field generated by the Hamiltonian $h_{\xi}=1$.  
{Interestingly, Legendre transformations correspond to discrete points along the orbits of   
the vector field $X_{\rm LT}$ generated by \cite{CesarDavid}
	\beq
	h_{\rm LT}=\frac{1}{2}\,{\sum_{a}\left[(q^{a})^{2}+(p_{a})^{2}\right]}.
	\eeq
{Furthermore,
	\beq
	\pounds_{X_{\rm LT}}\eta=0 \quad  \text{while} \quad \pounds_{X_{\rm LT}}G_{\rm FR} =- \sum_{i=1}^n \left(\d q^i \otimes \d q^i - \d p_i \otimes \d p_i\right),
	\eeq 
that is, infinitesimal Legendre transformations are symmetries}
of the contact structure but not of the metric structure of the TPS. 
Again, this is equivalent to saying that the First Law of thermodynamics
is invariant under Legendre transformations, while the fluctuations of the system's parameters change in the different ensembles.

The Jacobi bracket \eqref{Jacobibr} in general does not satisfy Leibniz rule and one has that $\{g,1\}_{\eta}=0$ if and only if $[X_{g},\xi]=0$. 
Nevertheless, if we restrict our attention to basic functions (resp. to strict contactomorphisms), then Leibniz rule is satisfied. 
When the 1-form $\eta$ defining the contact structure and the Hamiltonian function $h$ are fixed on $\mathcal T$,
we say that the quadruple $(\mathcal T,\mathcal D, \eta,h)$ is a \emph{Hamiltonian contact structure} or a \emph{contact Hamiltonian system}.
 We can express the action of $X_{h}$ on a function $f$ in terms of the Jacobi bracket \eqref{Jacobibr} as
	\beq\label{XhJacobi}
	X_{h}f=\xi(h)\,f +\{h,f\}_{\eta}.
	\eeq}

We say that a function $f\in C^{\infty}(\mathcal T)$ is a \emph{first integral} of the contact Hamiltonian structure $(\mathcal T,\mathcal D, \eta,h)$ if
$f$ is constant along the flow of $X_{h}$, that is if $X_{h}f=0$.
Notice that by equation \eqref{XhJacobi} this does not coincide with $\{h,f\}_{\eta}=0$, as in symplectic geometry.
From the above equation \eqref{XhJacobi} it follows that in general $X_{h}h=\xi(h)h$. Therefore the Hamiltonian function is not in general a first integral
of its flow. Indeed $h$ is a first integral if and only if it is a basic function. In this case we say that $h$ is a \emph{good Hamiltonian with respect to $\eta$}.}
Finally, given two first integrals $f_{1}$ and $f_{2}$ of the flow, we say that they are \emph{in involution} if $\{f_{1},f_{2}\}_{\eta}=0$ and we say that they are
\emph{independent} if their corresponding Hamiltonian vector fields $X_{f_1}$ and $X_{f_2}$ are linearly independent \cite{boyerintegrablesystems}.
Notice that equation \eqref{XhJacobi} can be used to construct invariant measures for non-conservative systems \cite{alediego}.


According to equation \eqref{generic.ham}, the flow of $X_{h}$ can be explicitly written in Darboux coordinates as
	\begin{empheq}[left=\empheqlbrace]{align}
	\label{z1}
	& \dot\w \,\,\,= h - p_a \frac{\partial h}{\partial p_a}\,,\\
	\label{z2}
	& \dot{p}_{a} \,\,=  -\frac{\partial h}{\partial q^a} + p_a \frac{\partial h}{\partial \w}\,,\\
	\label{z3}
	& \dot{q}^{a} =  \frac{\partial h }{\partial p_a} \,,
	\end{empheq}
where the dot denotes differentiation with respect to a parameter $t$ along the integral curves of $X_h$. The similarity with Hamilton's equations of classical mechanics is manifest. 
In fact, these are the contact equivalent to Hamilton's equations. In particular, when $h$ is a basic function
 equations \eqref{z2} and \eqref{z3} give exactly Hamilton's equations \cite{Arnold}.
Despite this similarity, there is a profound difference with Hamilton's equations because in general
$h$ is not conserved along the orbits of its own contact Hamiltonian field $X_h$. 

\subsection{The Thermodynamic Contact Hamiltonian System}
\subsubsection{Geometric properties}

Let us consider a curve on the TPS, $\gamma(t):I\subset \mathbb R\rightarrow \mathcal T$. If $\gamma(t)$ 
is an equilibrium process, then it must satisfy condition \eqref{newgammadot}. 
This means that projected on $\mathcal{E}$ 
it satisfies the First Law \eqref{w1law}. A coordinate expression for this condition is given by 

\beq\label{firstlawdot}
\dot\w=-p_a(\gamma)\dot{q}^{a}\,.
\eeq
Then, using equations \eqref{z1} and \eqref{z3} this implies that along the integral curves of $X_{h}$ that are constrained by the First Law - i.e. along equilibrium processes - 
we must have $h(t)\equiv 0$ {[c.f.  equations \eqref{newgammadot} and \eqref{isomorphismLiealg}]}. 
It follows that in the contact Hamiltonian formulation of thermodynamics we must have a Hamiltonian function that identically vanishes over all the orbits 
corresponding to equilibrium processes. We will now look for the most general form of such function.

{Let us mention that different candidates with such property can be found. To see that, 
consider the general expression for Legendre sub-manifolds as given in \eqref{LegsubArnold}.
From such expression there are some evident families of functions that vanish on the Legendre sub-manifolds.
Indeed Mrugala in \cite{mrugala2000} has studied three different families of such Hamiltonians, that is
\beq\label{Hmrugala}
h^{i}\equiv q^{i}-\frac{\partial f}{\partial p_{i}} \qquad h_{j}\equiv p_{j}+\frac{\partial f}{\partial q^{j}} \qquad h^{0}\equiv w-f + p_{i}\frac{\partial f}{\partial p_{i}}\,,
\eeq
where $f=f(p_{i},q^{j})$ as in \eqref{LegsubArnold}. However, due to the explicit presence of the thermodynamic fundamental relation $f=f(p_{i},q^{j})$
in these expressions, the flows involve $f$ and its derivatives and therefore they \emph{depend} on the particular choice of the system.
Here we wish to take a different route and define a contact Hamiltonian that could provide us with general information, valid for any system.}

 

{One of the basic assumptions in ordinary equilibrium thermodynamics (sometimes also listed as one of the postulates of thermodynamics)
is that the thermodynamic entropy $S_{\rm TD}$ for any system must be a homogeneous function of order one 
of the extensive variables \cite{Callen}. 
Therefore, 
{let us assume now that the $p_{a}$ are all the extensive variables, i.e. we fix the ensemble to be the $\mu$PT ensemble, according to section \ref{GibbsSM}}.
Since $S_{\rm TD}$ is a homogeneous function, that is
	\beq\label{HomPotential}
	S_{\rm TD}(\lambda p_{a})=\lambda S_{\rm TD}(p_{a}), \qquad \forall\lambda \in \mathbb R^{+}\,,
	\eeq
Euler's theorem for homogeneous functions implies
	\beq\label{EulersTh}
	S_{\rm TD}(p_{a})=p_aq^a,\qquad q^{a}=\frac{\partial S_{\rm TD}}{\partial p_{a}}.
	\eeq
Given the $p_{a}$ and the $q^{a}$, we may take the first equation in \eqref{EulersTh} as the definition
of the equilibrium entropy for any system. Moreover, as usual, we promote the $q^{a}$ to be independent of the $p_{a}$
when referring to functions on the TPS.
For instance, using standard thermodynamic coordinates, the first equation in \eqref{EulersTh} 
reads
	\beq\label{ETh2}
	S_{\rm TD}=\frac{1}{T}\,U+\frac{p}{T}\,V-\sum_{i=1}^{n-2}\frac{\mu_{i}}{T}\,N_{i},
	\eeq
where the intensive variables can be seen either as depending on the extensive ones, meaning that we are on the equilibrium 
manifold $\mathcal E$, or as independent coordinates, which means that we are considering $S_{\rm TD}$ as a function on $\mathcal T$. 

 The first equation in \eqref{EulersTh}
  serves as  a motivation to define the contact Hamiltonian function for thermodynamics $H:\mathcal T\rightarrow \mathbb R$ as follows
	\beq\label{ECH}
	H\equiv S_{\rm TD}-S_{\rho_{0}},
	\eeq
where $S_{\rho_{0}}$ is given by \eqref{macroEnt}.

Notice that, from the form of the macroscopic entropy  \eqref{macroEnt} and Euler's theorem \eqref{EulersTh}, 
it follows that
	\beq\label{HinT}
	H=-\w.
	\eeq
Moreover, in the thermodynamic limit -- i.e. suppressing fluctuations -- the statistical
entropy $S_{\rho_{0}}$ of Gibbs' distribution exactly coincides with the thermodynamic entropy $S_{\rm TD}$ for any system. 
This implies that $H$ vanishes on the equilibrium (Legendre) sub-manifolds of $\mathcal T$. 
{In fact, as discussed in section \ref{GibbsSM}, it follows from} equation \eqref{macroEnt} that $w$ is the total Legendre transform of the entropy, which is identically zero
for any extensive system at equilibrium. 
Therefore the information contained in $H$ will be the same for any system. In this sense
$H$ is a good candidate as a contact Hamiltonian function for thermodynamics.


The independence of $H$ of the fundamental relation is a major difference in our work from the approach in \cite{{mrugala2000,mrugalachina,mrugalaJPA}}.
Moreover, from the definition \eqref{ECH} we expect that $H$ is a good measure of how far the system is from equilibrium. In fact, on the one side 
the ensemble entropy $S_{\rho_{0}}$ and the thermodynamic entropy $S_{\rm TD}$ 
shall coincide at equilibrium and, therefore, $H$ must vanish. 
On the other side, $S_{\rho_0}$ subject to the constraints \eqref{constraint2}-\eqref{constraint1} 
reaches its maximum value $S_{\rm TD}$ when the system is in equilibrium. Thus for processes out of equilibrium we have that $S_{\rho_0} < S_{\rm TD}$.}
Therefore, for all spontaneous processes occurring on any system near equilibrium we must have
	\beq\label{Hgeq0}
	H\geq 0.
	\eeq
For this reason, $H$ is an appropriate contact Hamiltonian, both for its mathematical properties as well as for its physical meaning.

Let us see now  how to define equilibrium processes and admissible non-equilibrium processes by means of the flow
of the contact Hamiltonian vector field associated to $H$.
{According to \eqref{generic.ham}, the Hamiltonian vector field associated to \eqref{ECH} can be written in local Darboux coordinates as 
\beq\label{XE}
	X_{H} = -w\frac{\partial}{\partial \w}  - p_{a}\frac{\partial }{\partial p_{a}},
\eeq
which generates the homothety of $\eta$
\beq\label{XEhomothety}
\pounds_{X_{H}}\eta=-\eta.
\eeq
}
Notice that $X_{H}H=-H$ and therefore $H$ is not a first integral of its flow. 
Moreover, it follows from \eqref{XE} that the flow of $H$ reads
	\begin{empheq}[left=\empheqlbrace]{align}
	\label{z1ECH}
	& \dot\w \,\,\,= -\w\,,\\
	\label{z2ECH}
	& \dot{p}_{a} \,\,= - p_{a}\,,\\
	\label{z3ECH}
	& \dot{q}^{a}\,\, = 0 \,.
	\end{empheq}
Now let us first consider the geometrical properties of this flow as a contact Hamiltonian system. Then we will give it a meaning in the thermodynamic context.
It is immediate from \eqref{z3ECH} that the functions $q^{a}$ are $n$ first integrals of the flow. Moreover, the function $1$  provides another (trivial) first integral. 
Therefore, we have $n+1$ first integrals of the flow \eqref{z1ECH}-\eqref{z3ECH} and it is easy to check that they are in involution and independent.
This means that the contact Hamiltonian system
$\left(\mathcal T,\mathcal D,\eta,H,\{1,q^{a}\}\right)$
 is \emph{a completely integrable system of Reeb type} \cite{boyerintegrablesystems}.
 }
 
\subsubsection{Physical properties} 

Let us now turn to a more {physical} investigation of the integral curves of the flow
 $\gamma(t):I\subset\mathbb R\rightarrow \mathcal T$, which read
	\begin{empheq}[left=\empheqlbrace]{align}
	\label{z1int}
	& \w(t) \,\,\,= {\w}_{0}\,{\rm e}^{-t}\,,\\
	\label{z2int}
	& {p}_{a}(t) \,\,=  p_{a}^0\,{\rm e}^{-t}\,,\\
	\label{z3int}
	& {q}^{a}(t) \,\, = {q}^a_{0} \,.
	\end{empheq}
From \eqref{z1int}-\eqref{z3int} and the definition of $H$ -- equation \eqref{HinT} -- 
it follows that 
	\beq\label{h(t)}
	H(t)=H_{0}\,{\rm e}^{-t}.
	\eeq
 This means that we have two types of orbits for the flow:
 \begin{itemize}
 \item[i)] Orbits starting with $H_{0}=0$. Along these orbits $H(t)\equiv 0$ for all $t>0$.
 \item[ii)] Orbits starting with $H_{0}\neq0$. Along these orbits $H(t)$ tends exponentially to zero as $t$ increases.
 \end{itemize} 
  
  \begin{figure}
  \begin{center}
\includegraphics[width=0.75\columnwidth]{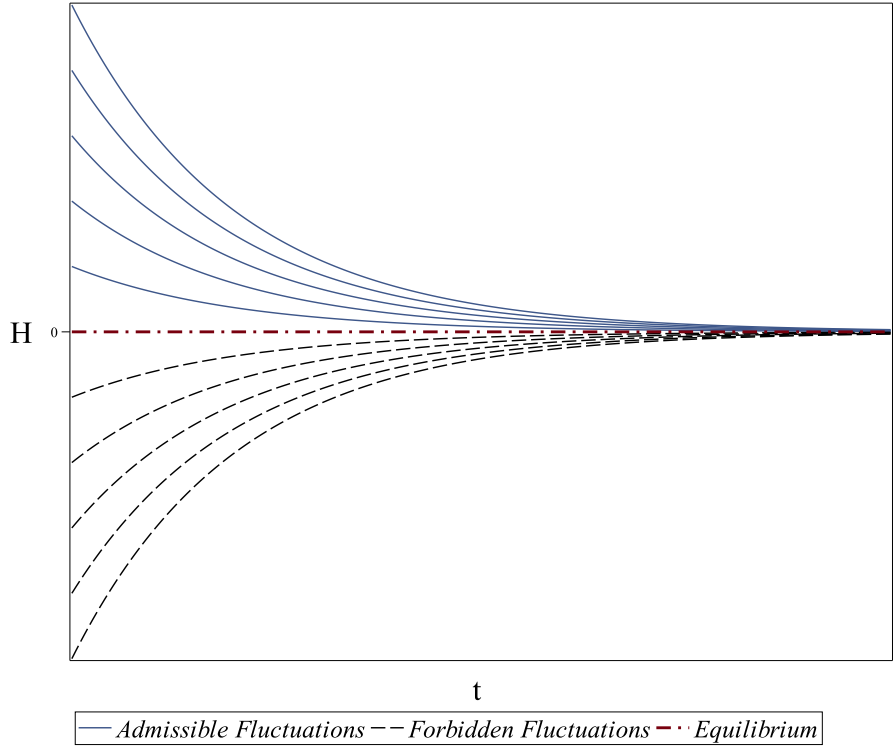}
\end{center}
\caption{Evolution of the irreversible entropy production $H(t)$ 
along the orbits of the thermodynamical flow. Solid lines represent admissible fluctuations. As we see, the entropy 
production along a fluctuation is a positive and monotonically decreasing function of $t$ which tends exponentially to zero.
We interpret this fact as a general feature representing geometrically in the TPS the evolution towards equilibrium.}
\label{figh(t)}
\end{figure}

{Recall that along equilibrium processes $H(t)$ must be zero by definition -- c.f. equation \eqref{ECH}.
To clarify the physical significance of $H$, let us show that
 the function $H$ here coincides with the irreversible entropy production necessary in order to re-establish equilibrium during a thermodynamic fluctuation.
 In fact, given a perturbation of the system out of the equilibrium value, the entropy change is
 	\beq\label{perturb}
	S=S_{\rm TD}+\delta S+\frac{1}{2}\delta^{2}S.
	\eeq
Considering that $\delta S=0$ at equilibrium, therefore 
	\beq\label{Hperturb}
	H=S_{\rm TD}-S=-\frac{1}{2}\delta^{2}S
	\eeq 
is the irreversible entropy
production in the linear regime, which is the Lyapunov function governing the damping of thermodynamic fluctuations \cite{NobelLecture}.

According to the above discussion, we now give the geometric definition of thermodynamic processes, both at equilibrium
and resulting from fluctuations.

In general, a \emph{thermodynamic process} is a orbit of the flow of $X_{H}$. 
 In particular, \emph{equilibrium (quasi-static) processes} are the orbits of the flow of $H$ of type i)  -- with $H\equiv 0$, i.e. no irreversible entropy production -- 
 while \emph{fluctuations} are
 orbits of type ii), since $H(t)$ is not zero but becomes negligibly small very rapidly.
 Moreover, a fluctuation is \emph{admissible} if and only if  $H_{0}=H(\gamma(0))\geq 0$ (i.e. the entropy production is positive).

As we see, with these definitions 
we can give a complete contact Hamiltonian characterization of equilibrium processes as well as of fluctuations and thermalization by means of the dynamics in the TPS.
This result sets a parallel between the phase space symplectic Hamiltonian description of conservative mechanics and
the phase space contact Hamiltonian description of thermodynamic processes, both equilibrium and fluctuations. 
}

\subsection{Thermodynamic length and entropy production}

Notice that we have given all the definitions of thermodynamic processes without using the metric structure of the TPS.
Now let us consider the metric $G_{\rm FR}$ on $\mathcal T$ as defined in \eqref{GinTcoordinate}.
A direct calculation shows that 
	\beq\label{normsquared}
	||\dot \gamma(t)||^{2}=G_{\rm FR}(X_{H},X_{H})=H(t)^{2}
	\eeq
and therefore
the square norm of such processes is always positive.
This means that the metric structure does not distinguish between admissible and non-admissible thermodynamic processes. 
 
 Nevertheless, it is remarkable that the square norm is exactly the square of the thermodynamic Hamiltonian. In fact, this enables us to 
 define a functional which vanishes for equilibrium processes and allows us to calculate the entropy production for any admissible near-equilibrium process.
 Given the pseudo-Riemannian structure $G_{\rm FR}$ and the fact that $X_{H}$ has a non-negative squared norm -- c.f. \eqref{normsquared} --, 
  we use the \emph{arc-length functional}  
 	\begin{equation}\label{arclength}
	\begin{split}
	&{\mathcal S}:\,\,\mathcal C(\mathcal T) \,\,\,	\rightarrow \,\,\mathbb R\\
	&\gamma(t) \,\,\,\,\mapsto \int_{0}^{t_{f}}\sqrt{G_{\rm FR}(X_{H},X_{H})}\,\d t\,.
	\end{split}
	\end{equation}
{where $\mathcal{C}(\mathcal{T})$ denotes the space of differentiable functions on the TPS.} 
Then, $\mathcal S$  is the \emph{total irreversible entropy produced along a process}.
From \eqref{normsquared} and \eqref{arclength} it follows that {along equilibrium processes the total entropy production
thus defined is zero, whereas for  fluctuations} we have
	\beq
	{\mathcal S}(\gamma(t))=\int_{0}^{t_{f}}H(t)\,\d t=\int_{0}^{t_{f}}H_{0}{\rm e^{-t}}\,\d t=H_{0}\left[1-{\rm e}^{-t_{f}}\right]\geq0\,.
	\eeq
Noticeably ${\mathcal S}(\gamma(t))$ has a global {minimum} $\mathcal S=0$, which is attained if and only if $S(\gamma(t))\equiv 0$, that is if the process is of type i), i.e. if and only
if $\gamma(t)$ is an equilibrium process for which the initial condition is $H_0 =0$.
Hence, we conclude with the result that equilibrium processes are those paths for which the total entropy production is identically zero. 
{Notice finally that the total entropy production for any fluctuation when $t_{f}\mapsto \infty$ is exactly $H_{0}$. Using 
the relation \eqref{Hperturb} we see that our definition coincides with the expectation from ordinary thermodynamics, i.e. that the system
relaxes to equilibrium after producing an amount of entropy corresponding to the initial displacement from the equilibrium entropy.
In this way we prove that the total length of any thermodynamic process equals its entropy production.}

\section{Discussion of the results}\label{sec:conclusions}
Let us summarize and discuss the new results presented in this work. 
To the best of our knowledge, the role of contact gauge transformations of the Thermodynamic Phase Space had never been considered before. 
These transformations were presented here in their full generality in \ref{sec:gaugetransf}. We have then considered the particular case of the change of 
the thermodynamic representation, showing explicitly the  transformation of the {para-}Sasakian structure of the TPS. 
Interestingly, such change in the metric 
structure induces the well-known conformal equivalence between Weinhold's `energy' metric and Ruppeiner's `entropy' metric on the Legendre 
sub-manifolds. 
It also reveals interesting scaling properties of the almost para-contact structure, which may provide us with further insights 
 on its physical interpretation. This aspect has not been considered here and it will deserve attention in future works \cite{usinpreparation}.

In section \ref{sec:Lsym} we have investigated in detail the role of the Legendre symmetry in thermodynamics. We have proved that this symmetry induces
a diffeomorphism of the equilibrium sub-manifolds and therefore implies ensemble equivalence, as long as the stability conditions are fulfilled.
Moreover, we have also proved that a total Legendre transformation is an isometry {between the Hessian} metric structures naturally induced by {the Fisher Information Matrix} 
on the Legendre 
sub-manifolds by the use of different embeddings, while a partial Legendre transformation is not so. 
In this respect, it is important to note that our approach differs from previous literature  on thermodynamic geometry. In fact, in the literature one usually starts directly
 with a particular choice of a metric on the Legendre sub-manifold and then operates on it with a change of the coordinates of the sub-manifold which leave the metric
unchanged. In our case,  since the TPS is locally isomorphic to the hyperbolic Heisenberg group, its metric $G_{\rm FR}$ is fixed. Thus, we consider the different metrics that can be induced on its Legendre 
sub-manifolds.  
We have also presented a digression on first order phase
transition as a relevant example of regions of ensemble inequivalence in the context of ordinary thermodynamics {and fixed their geometric representation in the TPS.}

Finally, in section \ref{sec:ECH} we have given a consistent formulation of thermodynamic processes in terms of a dynamical system on the contact phase space.
We have shown that {the irreversible entropy production in the local equilibrium regime} is a good contact Hamiltonian, 
in the sense that it defines a flow which is completely integrable and whose 
integral curves represent thermodynamic processes, both at equilibrium and 
 near-equilibrium. An interesting result here is that 
 we can prove by means of the contact dynamics at the level of the phase space of any thermodynamic system that thermodynamic fluctuations
 vanish.
Such formulation of equilibrium processes and thermodynamic fluctuations places the contact Hamiltonian description of thermodynamics
on an equal footing as the symplectic Hamiltonian description of conservative mechanics. 
This parallelism suggests that contact dynamics could be a good candidate for the description of
the statistical mechanics of non-conservative systems, as pointed out in \cite{alediego} and it is certainly worth of further investigation. 
 Finally, we have also shown that the metric structure $G_{\rm FR}$
can be exploited to compute the entropy production of any process by means of the arc-length.

It is worth noticing at this point that the standard problem of equilibrium thermodynamics is that of determining the final state of the evolution of a system placed
in contact with external reservoirs. In fact, the resolution to this problem is given by the \emph{extremum principle}, which can be formulated in many equivalent ways,
depending on the thermodynamic potential considered. In particular, we have shown that the contact Hamiltonian evolution
reproduces the evolution predicted by the `Entropy Maximum Principle' as follows. 
\begin{itemize}
\item{\bf Entropy Maximum Principle}.
The equilibrium value of any unconstrained internal parameter
in a system in contact with a set of reservoirs ({with intensive parameters $q^{1}_{r},\dots, q^{n}_{r}$}) {maximizes the thermodynamic entropy $S(p_{1},\dots,p_{n})$
at constant $q^{1},\dots,q^{n}$} ({equal to $q^{1}_{r},\dots,q^{n}_{r}$}) \cite{Callen}.

The evolution that we have found from contact Hamiltonian dynamics -- equations \eqref{z1ECH}-\eqref{z3ECH} -- 
exactly matches this entropy maximum principle, i.e. the intensive variables $q^{a}$ have fixed values (equal to those of the external reservoirs), while the internal extensive parameters
$p_{a}$ evolve so as to find the maximum value of the entropy for the given constraints.
This process requires an entropy production which is characterized by the evolution of $H=-w$ and whose total amount is exactly the same
quantity predicted in thermodynamics for a fluctuation.
\end{itemize}

Finally, let us resume schematically  the results of section \ref{sec:ECH}, i.e. the contact Hamiltonian formulation of thermodynamics and its connection with the classical Laws of thermodynamics. 

\begin{itemize}
\item {\bf Zeroth Law}. 
	Notice that we are always working on a specific ensemble, obtained from maximizing Gibbs' entropy functional subject
	to some `boundary conditions'. Therefore the equilibrium condition is assumed because of the use of the Gibbs distribution corresponding to the ensemble.
	Accordingly, it turns out that the contact Hamiltonian flow has the intensive parameters $q^{a}$ as \emph{first integrals of motion}, which is in agreement 
	with the evolution of a macroscopic system in contact with temperature and particle reservoirs predicted by phenomenological thermodynamics.

\item {\bf First Law}. 
	Our definition of equilibrium processes -- integral curves of the flow of the function $H$ for which $H\equiv 0$ -- automatically satisfies the First Law
	of thermodynamics, c.f. equations \eqref{ECH} and \eqref{z1ECH}-\eqref{z3ECH}.
	Moreover, by the definition of $H$ -- equation \eqref{ECH} -- this 
	is equivalent to say that there is no irreversible entropy production along the process.
	
\item {\bf Second Law}. 
	The Second Law of thermodynamics establishes that entropy is a maximum at equilibrium and that  entropy production along a spontaneous fluctuation
	is non-negative.
	This provided us with a geometric definition of \emph{thermodynamically admissible processes}, i.e. processes for which $H_{0}\geq 0$.
	By the flow of $H$ this also implies
	that entropy production is non-negative along all the process and that fluctuations vanish and the system thermalizes exponentially in the flow parameter $t$.\\
\end{itemize}

{To conclude, in this work we have presented a thorough analysis of some open problems in the geometrization of thermodynamics.
In particular, here we have focused on gauge transformations, Legendre symmetries, first order phase transitions and on the construction of a comprehensive contact Hamiltonian system
that entails the Laws of thermodynamics from a geometric perspective over the phase space.
Nevertheless, several questions remain to be addressed. For example, we have not fully investigated the physical role of gauge transformations and
of the other symmetries of the contact structure corresponding to strict contactomorphisms which are not Legendre. From the point of view of contact Hamiltonian thermodynamics, 
it would be relevant to see whether this formulation can lead e.g. to a kind of  
quantization of the contact manifold and of the 
thermodynamic relations, as previously proposed e.g. in \cite{rajeevquantization}. Moreover, we understand that the length of a process in the phase space is related 
to the irreversible entropy production during the process. Therefore it is interesting to perform a detailed analysis of the geodesics, as curves of minimal entropy production, as in \cite{SalamonBerryPRL}
and \cite{Crooks,CrooksPRE2012,CrooksPRL2012}.
We expect to explore all these topics in future works.}




\section*{Acknowledgements}
{The authors are thankful to M. A. Garcia Ariza and H. Quevedo for insightful comments and suggestions.}
AB wants to express his gratitude to the A. Della Riccia Foundation (Florence, Italy) for financial support. 
CSLM was supported by DGAPA-UNAM (postdoctoral fellowship). 
FN  acknowledges financial support from CONACYT grant No. 207934.

\end{document}